\documentclass[lettersize,journal]{IEEEtran}
\usepackage{amsmath,amsfonts}
\usepackage{algorithmic}
\usepackage{algorithm}
\usepackage{array}
\usepackage[T1]{fontenc}
\usepackage{textcomp}
\usepackage{stfloats}
\usepackage{url}
\usepackage{verbatim}
\usepackage{graphicx}
\usepackage{cite}
\usepackage[sort, numbers]{natbib}
\usepackage[hidelinks]{hyperref}
\usepackage{color,soul}
\usepackage{tikz,graphics,color,fullpage,float,epsf,caption,subcaption}
\usepackage{authblk}

\begin{document}

\title{MITHOS: Interactive Mixed Reality Training to Support Professional Socio-Emotional Interactions at Schools}

\author[1]{Lara Chehayeb}
\author[1]{Chirag Bhuvaneshwara\thanks{firstName.lastName@dfki.de}}
\author[1]{Manuel Anglet}
\author[1]{Bernhard Hilpert}
\author[1]{Ann-Kristin Meyer}
\author[1]{Dimitra Tsovaltzi}
\author[1]{Patrick Gebhard}
\author[2]{Antje Biermann\thanks{a.biermann@mx.uni-saarland.de}}
\author[2]{Sinah Auchtor\thanks{sinah.auchtor@uni-saarland.de}}
\author[3]{Nils Lauinger\thanks{nils.lauinger@fobid.org}}
\author[3]{Julia Knopf\thanks{julia.knopf@mx.uni-saarland.de}}
\author[4]{Andreas Kaiser\thanks{firstName.lastName@tricat.net}}
\author[4]{Fabian Kersting}
\author[4]{Gregor Mehlmann}
\author[5]{Florian Lingenfelser\thanks{lastName@hcm-lab.de}}
\author[5]{Elisabeth André}

\affil[1]{DFKI GmbH, Kaiserslautern}
\affil[2]{Saarland University, Saarbrücken}
\affil[3]{FoBiD, Saarbrücken}
\affil[4]{TriCat GmbH, Ulm}
\affil[5]{University of Augsburg, Augsburg}




\maketitle

\begin{abstract}
Teachers in challenging conflict situations often experience shame and self-blame which relate to the feeling of incompetence \cite{hargreaves2000mixed}, but may externalise as anger \cite{nathanson2014affect}. Sending mixed signals fails the contingency rule for developing affect regulation \cite{gergely1996social}, and may result in confusion for students about their own emotions and hinder their emotion regulation. Therefore, being able to constructively regulate emotions not only benefits individual experience of emotions but also fosters effective interpersonal emotion regulation, and influences how a situation is managed \cite{frenzel2021teacher}. 
MITHOS is a system aimed at training teachers’ conflict resolution skills through realistic situative learning opportunities during classroom conflicts. In four stages, MITHOS supports teachers’ socio-emotional self-awareness, perspective taking and positive regard. It provides a) a safe virtual environment to train free social interaction and receive natural social feedback from reciprocal student-agent reactions, b) spatial situational perspective taking through an avatar c) individual virtual reflection guidance on emotional experiences through co-regulation processes and d) expert feedback on professional behavioural strategies. This chapter presents the four stages and their implementation in a semi-automatic a Wizard-of-Oz (WoZ) System. 
The WoZ system affords collecting data that are used for developing the fully automated hybrid (machine learning and model-based) system, and to validate the underlying psychological and conflict resolution models
We present results validating the approach in terms of scenario realism, as well as of a systematic testing of the effects of external avatar similarity on antecedents of self-awareness with behaviour similarity. The chapter contributes to a common methodology of conducting interdisciplinary research for human-centered and generalisable XR and presents a system to designed support it. 
\end{abstract}


\section{Introduction}
\IEEEPARstart{I}{n} social interactions, people learn from each other through situative experiences. How others react to our behavior can provide rich situative social feedback about what needs to be adjusted. Emotions comprise important indicators on how to assess a situation and which options for action are available to us \cite{gross2014emotion}. Teachers, similar to parents, hold a central role as multipliers of socio-emotional competence and behavior thereof, which takes place largely also through unaware socialisation processes \cite{Dale2015, eckstein2015adventure}. However, teachers often feel inadequately prepared to deal with complex socio-emotional challenges, such as those that can be triggered by heterogeneous learning groups or behavioral conflicts. The goal is to strengthen teachers’ professional communication and empathic competence. To this end, effective methods for emotion regulation and conflict resolution are required \cite{gross2014emotion, keller2001konfliktmanagement}, which cannot be trained systematically and realistically in education. \emph{MITHOS, an Interactive Mixed Reality Training to Support Professional Socio-Emotional Interactions at Schools}, examines the potential of an interactive mixed reality training to support the ability of teachers to emotionally assess conflict situations in the classroom and reevaluate possibilities for behaving with the students. 

Students are sharp in perceiving subtle emotions and underlying aggression in teacher behavior, and recursive teacher-student behavioral influences disrupt classroom interaction \cite{frenzel2021teacher}, and result in conflict escalation. Emotion regulation, like self-compassion, and perspective taking in learning contexts can support understanding and coping with one’s own and others´ emotions \cite{park2022implicit}. Training teachers’ emotion regulation and perspective taking can thus improve subtle (non-)verbal behavior and conflict regulation. However, teachers don’t have opportunities to be trained on realistic conflict situations and mostly have theoretical knowledge, before they teach in real classrooms \cite{krause2013arbeitssituation}, leaving teachers with no available personal socio-emotional resources. The MITHOS training system supports the acquisition of professional social skills for teaching. MITHOS is a multi-modal Mixed Reality (MR)-Training for teachers to enhance their conflict regulation strategies. It employs realistic situative learning opportunities during challenging classroom social situations. Socially and emotionally challenging situations are realized in a virtual-immersive MR environment. In this environment, a human-controlled avatar can communicate with multiple autonomous agents representing students. In this way, the ability to empathize and deal professionally with conflict situations in the classroom is trained and alternative courses of action are demonstrated for practical use. The training focuses on perspective taking and conflict regulation. 

The relative realism but reduced complexity of the training situation helps to bridge the gap between theory and classroom. Teachers interact in a safe virtual classroom environment with socially interactive student-agents (SIA) \cite{10.1145/3477322} whose behavior induces realistic conflict situations. The conflict interaction is implemented in an automatic modeling process and the SIA student’s conflict behavior is generated. This provides immediate implicit and situative social feedback \cite{fonagym}, much like in human-to-human interaction. 


In an immersive MR interaction simulation (Training Stage 1), teachers may interact freely and experience different student reactions to their behavior, as the conflict evolves in several turns \cite{hartmann2023imagine}. Through the interaction, the teacher becomes more awareness of the conflict through an escalation of the behaviour. Subsequent training stages, support emotional self-awareness \cite{gyurak2011explicit} and empathic positive regard \cite{lenske2015linzer} through spacial perspective taking \cite{kessler2010embodied} and emotion co-regulation \cite{gergely1996social} to allow and integrated experience that can be put into action \cite{fonagyaffect}. The system plays back the teacher’s behavior using a virtual agent to simulate spacial perspective taking as a means to support emotional perspective taking \cite{batson1997perspective}. The second phase hence promotes further self-awareness in the form of reflection on the teacher’s own and the student’s emotional experiences and behavior \cite{hooi2014avatar}. The stages that follow the interaction serve to facilitate the teacher's successful experiences in further interactions, paving the way for transformation to occur \cite{moser1996entwicklung}


\section{Human-Centered System Development}

The human-centered and generalizable concept includes a novel combination of virtual-immersive and collaborative 3D environments, social signal analysis and synthesis of verbal, non-verbal and paraverbal behaviors to simulate interactive virtual agents. Together, this allows for realistic training. MITHOS, as a first project, enables the real-time exchange of social signals between humans and agents in a teaching setting, including natural implicit social feedback, as well as the coupling of virtual and real-world experiences. It thus simulates socio-emotional co-regulation. 

The implementation is done in two versions: 1) Wizard of Oz (WoZ) System (\ref{fig_Woz_sys_overview}), and 2) Automated System. The WoZ (Wizard of Oz) system is designed to empirically test and validate separate aspects of the behavior modeling and to collect data for developing the Automated System. The collected multimodal audio-visual data is used to develop a hybrid affect model, which utilizes both machine learning and theory-based reasoning models to decipher the situational emotional experience. This model is designed to process speech, gaze, pose, facial expression and natural language dialog, represented as an Interaction Act, that is a representation structure defined and used to understand the teacher’s behavior and emotional experience using computational models of affect, namely ALMA \cite{gebhard2005alma}
and structured psychodynamic diagnostic models \cite{opd2001operationalized}. The planned affect model then automatically generates appropriate student responses over multiple dialog turns. In the current WoZ system, the analysis is performed in real-time by a wizard trained to evaluate the teacher's behavior and the evaluation is automatically converted into suitable multimodal animation commands by the system. 
The main advantage of using psychodynamic models and deep modeling for co-regulation training lies in the long-term effectiveness for well-being and happiness, which goes beyond the treatment of symptoms. In this regard, psychodynamic methods are more suitable for interaction modeling, which does not target specific symptoms but generally observable, but relatively stable motives on a personal level, as described in the OPD (Operational Psychodynamic Diagnostic)\cite{pilling2020long}.

The MITHOS training, afforded by both the WoZ and the final Automated system, is anchored in the teacher training university programme of studies for systematic empirical investigation. The training situations can be run through repeatedly, and do not risk disadvantages for individuals. Both student-teachers  and experienced teachers can use these trainings to prepare purposefully for the challenges of teaching groups of learners who may have different cultural, emotional, and cognitive backgrounds.

\section{XR in Learning, Teacher Education, Conflict and its Resolution}
Current research in the field of virtual 3D learning and simulation environments has shown that the use of this technology can have great potential for learning and training effects \cite{greenwald2017technology,lerner2020immersive}. In teacher training, for example, MR is used to offer prospective teachers opportunities to practice strategic situational action in a protected environment \cite{dalinger2020mixed}. It has been shown that 3D learning and simulation environments with virtual interactive agents have high potential for social and emotional learning. Fields of application include preparing for lectures with the help of a virtual audience \cite{batrinca},

Previous MR approaches in conflict training \cite{dalinger2020mixed}, in which student reactions are taken from actors and are scripted, show advantages for authentic experiences, but have problems with credibility and personalized training. The perception of one's own needs and emotional arousal play a very important role here \cite{bautista2015exploring}. In addition to AI methods, group awareness tools (GATs) combined with behavioral support can highlight communication characteristics, support group processes and  perspective taking and interaction in learning contexts \cite{tsovaltzi2014group, puhl2015long, tsovaltzi2017leveraging}.

In the training of teachers, knowledge about social interaction \cite{baumert2011kompetenzmodell} and the perception and interpretation of cognitive and emotional processes \cite{praetorius2018generic} are particularly important. In order to successfully train these skills, an adequate theory-to-practice connection is essential. (Video-based) case vignettes are often used for corresponding training \cite{thiel2017}, but also virtual training \cite{kervin2006classsim, lugrin2016breaking}. Another important aspect of training is feedback \cite{hattie2009black}, which usefully consists of an immediate and a delayed component \cite{smith1993instructional}. This allows teachers to learn to decide quickly and flexibly on a reaction (immediate feedback) as well as to rethink complex conditions within a challenging situation \cite{lotz2015hattie}.

While there are previous existing projects focusing on providing a safe learning environment, their focus is mostly on training knowledge acquisition and teaching strategies \cite{Dieker2015TLETU,dieker2016mixed, CONNOLLY2012661}. MITHOS stands out as it focuses on training socio-emotional skills and classroom management. MITHOS emphasizes the importance of respectful and professional interactions in a classroom as a basis regardless of the actual teaching strategy or content used. It also emphasised teacher well-being. 

\section{Situational Interaction for Training Professional Respectful Conflict Resolution}
\subsection{Definition of Respectful Social Interaction}
Acquiring socio-emotional skills influences communication skills and professional behavior \cite{anuvziene2015structure} and contributes to health and well-being \cite{chang2009appraisal, keller2014teachers}. During social interactions, reactions are determined based on the appraisal of perceived emotional reactions of others \cite{manstead2001social}. This may increase their awareness of their behavior during high arousal situations, i.e. the initiated conflict. Self-awareness brings attention to not only one’s own emotions and needs, but also that of the others. By increasing self-awareness, a person is more likely to adopt adaptive regulating strategies \cite{Gross2015}. 

Perspective taking also enhances one’s ability to empathize \cite{ames2008taking}. High arousal levels may reduce the ability of perspective taking and trigger the switch of emotion regulation to automatic response through fight or flight. This switch may be influenced by different factors such as previous internal conflicts \cite{fonagy2006mechanisms}. In MITHOS, we induce self-focused attention through the replay of the interaction on an avatar to foster self-awareness and perspective taking (Training Stage 2). Based on previous empirical research \cite{scaffidi2016self} to ensure fostering perspective taking a cue to the other's perspective has to be present. In MITHOS, teachers watch a replay of the interaction from the perspective of the student, watching their own expressions on an avatar. 
Previous studies have shown that using a generic avatar is not sufficient and a personalized avatar is a prerequisite for increasing self-awareness \cite{10.1145/1240624.1240696}. Personalized avatars are used as they increase the feeling of being self-represented and emotionally attached to the visually similar avatar stimulus. 

In a previous study, we  investigated the effect of different degrees of similarity and the necessity to include expensive similar avatars for perspective taking in MITHOS. We conducted a study defining three conditions by manipulating specific facial features of the virtual avatar \cite{alves2023visual}. The results show that perceived similarity is influenced by two degrees of manipulations of avatar-person similarity. This allows some room for error in studies when designing personalized avatars. However, the right degree of manipulation is required for explicit identification with the avatar. Therefore, here we further investigate the added value of personalized avatars for increasing self-awareness. We  compare two different conditions, watching the replay of one's behaviour on a generic default avatar compared to a personalized avatar of maximum possible similarity using MetaHuman \footnote{\url{https://www.unrealengine.com/en-US/metahuman}.}, that can be used freely in the Unreal Engine \footnote{ \url{https://www.unrealengine.com/en-US}.} environment. 

Furthermore, perspective taking and self-awareness are influenced by one's internal conflict. Internal conflicts stem from experiences during development and are triggered by present interactions. Internal conflicts are experienced when one’s desires or socio-emotional needs are not met. Therefore, the internal conflict forms one's behavior to reach the desired state in the interaction \cite{https://doi.org/10.1017/S0048577201393198}. The behavior is additionally shaped by the situation and generally accepted social norms \cite{AJZEN1991179}. In MITHOS, we aim to track the lead affect of the teacher during the automated interaction, based on the OPD method\cite{opd2001operationalized}. Following the diagnostic model, we analyze the connection between the self-reported internal conflict of the teacher, the lead affect and the intentions based on qualitative analysis of an interview where teachers reflect on their experience during the interaction with the help of a trained expert interviewer. In the final system, we make use of this data, to automate the modelling of the situation-based teacher lead affect and intentions, and to simulate an empathic feedback agent to train socio-emotional skills, specifically self-compassion. 
  
Self-compassion improves emotion regulation \cite{NEFF2007139}, increasing the capacity to co-regulate. Training on self-compassion specifically may deactivate the threat system (associated with feelings of insecurity, defensiveness) and activate the self-soothing system \cite{gilbert1989human}. By training teachers to be self-compassionate, we influence their ability to self-regulate and co-regulate students, and in turn influence the student’s ability to self-regulate their emotions \cite{mainhard2018student}. In MITHOS finaly system, the teacher is supported by a training agent to reflect on the experience and provide support on empathic conflict regulation. The feedback is specific to the personalized experience and responses as detected by the automated system. In an attempt to target the affect regulation system, the agent trains compassion to self as socio-emotional skill. Enhancing the self-soothing system allows the teacher to step back from the situation and perceive it objectively with consideration to both oneself and the other. 

\subsection{Professional Conflict Regulation in the classroom (BiWI)}
Conflict regulation in the classroom – especially handling disturbances while teaching – is a crucial professional competence of teachers \cite{kunter2013development}. For novice teachers, dealing with conflicts in the classroom is a stress-related factor and they often don’t feel well prepared \cite{dicke2015reducing}. To deal with disturbances and conflicts, the three dimensions control (rules and routines, behavior control etc.), student-teacher relationship (positive regard, student participation etc.), and task focus (e.g. relevance of learning aims, cognitive activation) should be implemented in a balanced way \cite{lenske2015linzer, marzano2005handbook}. A positive student-teacher relationship includes authenticity, positive regard, empathic understanding, and positive emotionality. There is a positive relationship between positive regard and empathic understanding of a teacher with less aggressive or disturbing behavior from students, but also higher thinking levels and learning success in students \cite{hattie2008visible, Tausch2008PersonzentriertesVV, aspy1972investigation, aspy1974humane}. The effects can be explained with the behaviors of more empathic and compassionate teachers: For example, they tend to use more eye contact and smile more, and are more interested in their students’ feelings, and express appreciation toward them \cite{aspy1972investigation, aspy1974humane}. Despite the fact that positive regard, and empathetic understanding of students is an important basis for conflict regulation, the task focus and control dimensions should not be overlooked. Students should be motivated and activated to participate in the lesson (cognitive activation, \cite{kunter2013development}); the teacher should also manage the behavior of the students (behavior control), without neglecting the basic psychology needs of autonomy (cf. self-determination theory; \cite{ryan2017self}). In MITHOS, we want to evolve teachers’ ability to react in a balanced way to conflict behavior of students, that means to focus on the task and simultaneously on a positive relationship to the student. Respectively, the virtual students’ reactions are composed of two parts of addressing the relationship and the task dimension, to enable differentiated training. 

\subsection{Conflict situations in a heterogeneous classroom setting}
The debate on heterogeneity in schools is primarily focused on academic performance and achievement, which is due to the fact that academic performance is often used as a measure of success in the education system \cite{glock2020einstellungen}. 
Accordingly, the reactions of the virtual student are defined based on the representaion of conflict situation. Two  dimensions of heterogeneity are additionally implemented: the linguistic heterogeneity and the visual representation of the virtual student. These address stereotypes that teachers may hold against combinations of representations of those dimensions. Teachers may be affected by their stereotypes in the conflict and more likely decide to use more drastic methods \cite{nichols2004exploration, rocque2011understanding, skiba2002color, townsend2000disproportionate}.

The conflict behavior shown by the virtual student is thus based designed by expert teachers and a therapist who was responsible for the consistency of the student behaviour and the represented internal conflict. They all members of the research group. The scenario is composed of four conflict stages \cite{keller2001konfliktmanagement, darnon2006mastery} that are each structured in seven levels of conflict for the task and relationship part, mentioned in the previous chapter \cite{schwarz2013konfliktmanagement, rattay2011funf}. Therefore, a mapping between the conflict stages and the conflict levels for task and relationship was created and filled with spoken text samples, gestures and body movements (see chapter 3). As the conflict levels for task and relationship vary from each other, the behavior was created separately and the combinations were validated in a separate pilot study. In total this makes a set of 86 different behaviors composed of a verbal and a non-verbal part for one scenario. Currently two scenarios are scripted, based on the experiences made in a \textit{gymnasium} and a \textit{technical school (realschule)}, both in Germany. 
An analysis of the presence of academic language \cite{nagy2012words} in the two different scenarios is currently in progress. The analysis is oriented on the three dimensions of academic language: the lexical, grammatical and discursive features \cite{heppt2016verstandnis}. The way that the text files are spoken will also be variable: Two students (male and female) are recorded for the audio files, which are then rendered by an AI-voice model (https://elevenlabs.io/). The original sample files of the students are created in Saarland dialect and in standard German. This gives again the possibility to address another dimension of linguistic heterogeneity.

The linguistic heterogeneity, the gender and the skin color of the virtual student are randomized to not reinforce prejudices that may exist in form of combinations of the dimensions of the language and the visual representation of the virtual student.

\section{Interaction Simulation Stage}
The MITHOS-Training, whether WoZ system or Final Automated System, takes place in four main training Stages: 1. First situative training stage, 2. Avatar-replay stage, 3. Agent-feedback stage and 4. Expert-feedback stage.
All stages together aim at increasing the teacher’s awareness and ability of perspective taking, and training social interaction with a conflict scenario in an interactive classroom.

In a situative training stage, teachers train social interaction with an interactive classroom conflict scenario. They hold a class in front of a virtual classroom where a conflict is initiated through the behavior of one of the virtual students. The teacher is required to involve the student in the lesson and guide them to pay attention, in a professional and respectful way. The virtual student reactions are implemented reciprocally and are multimodal, to reflect different communication channels that include verbal and non-verbal, as well as explicit and implicit signals in the interaction. 
Through the signals, the teacher’s implicit emotional awareness increases. Implicit emotional awareness is non-effortful attention to one’s emotional experiences which directly influences one’s emotion regulation and in turn the conflict potential in the interaction. The next Stages provide feedback (implicit and explicit) and train the teacher to be more professional. In the avatar-replay stage of the interaction, teachers use an explicit emotion awareness tool, implemented as a virtual agent who plays back their verbal and non-verbal behavior during the interaction. This allows them to take the physical perspective of the students and observe their reactions, to increase their awareness of the student’s emotional perspective. In the agent-feedback stage, the teacher reflects on their experience with the help of  a virtual training agent in a process of co-regulation. The agent supports, through compassionate reactions to the emotional experience of the teacher. Compassion complements explicit emotional awareness \cite{Gyurak2011} as it reduces personal distress \cite{Duarte2016}. Finally, in the expert-feedback stage, the teacher gets verbal feedback from experts on positive regard as an important aspect on professional conflict regulation in the classroom.


\subsection{Training stage 1 $<$\textbf{Situative training stage}$>$} 
\begin{figure*}[htbp]
\centering
\includegraphics[width=\textwidth]{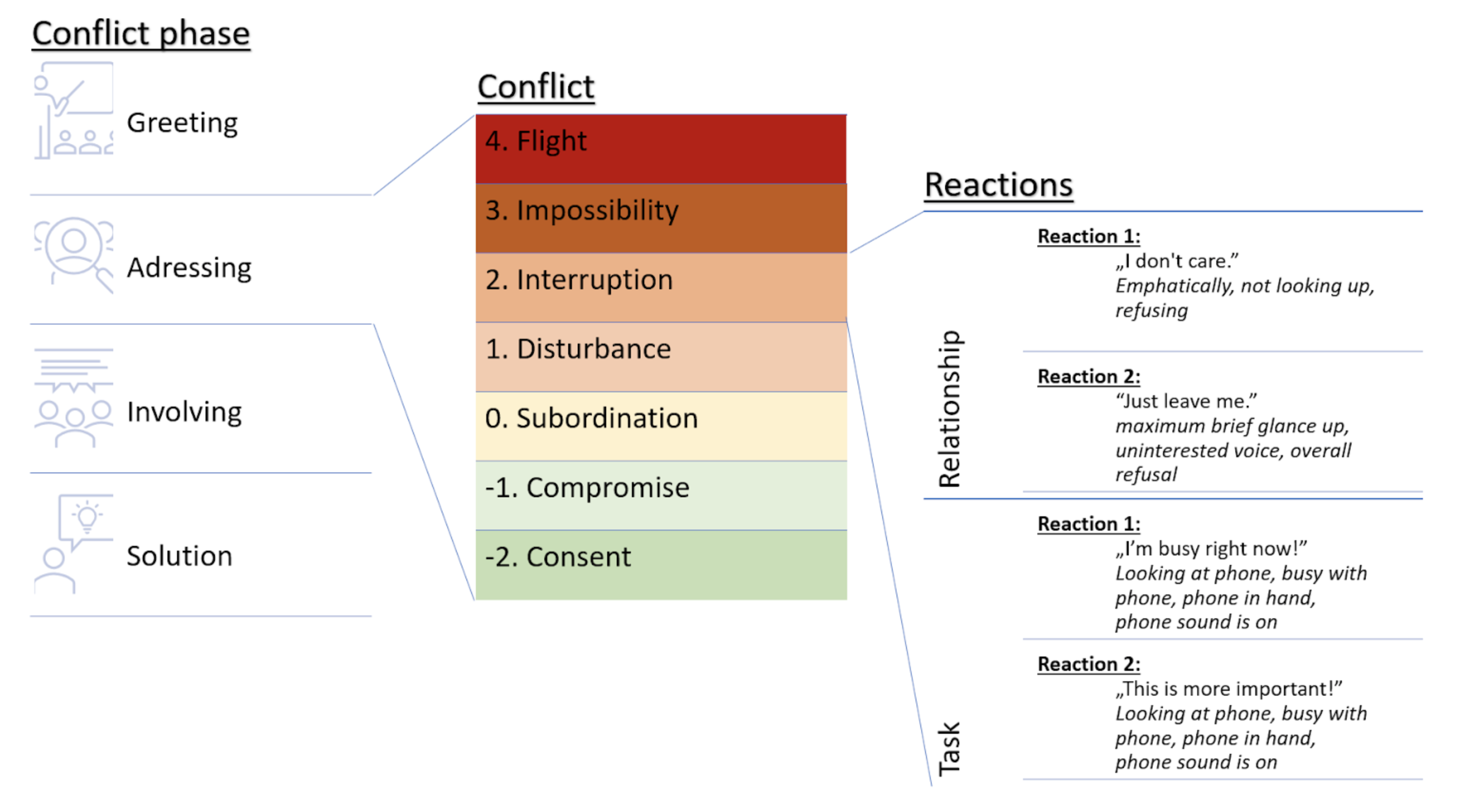}
\caption{Semi-automatic modeling of teacher behavior to be used for turn-by-turn evaluation of the teacher behavior on two dimensions: task and relationship conflict levels. }
\label{fig_teacher_behavior_model}
\end{figure*}

\begin{figure*}[htbp]
\centering
\includegraphics[width=\textwidth]{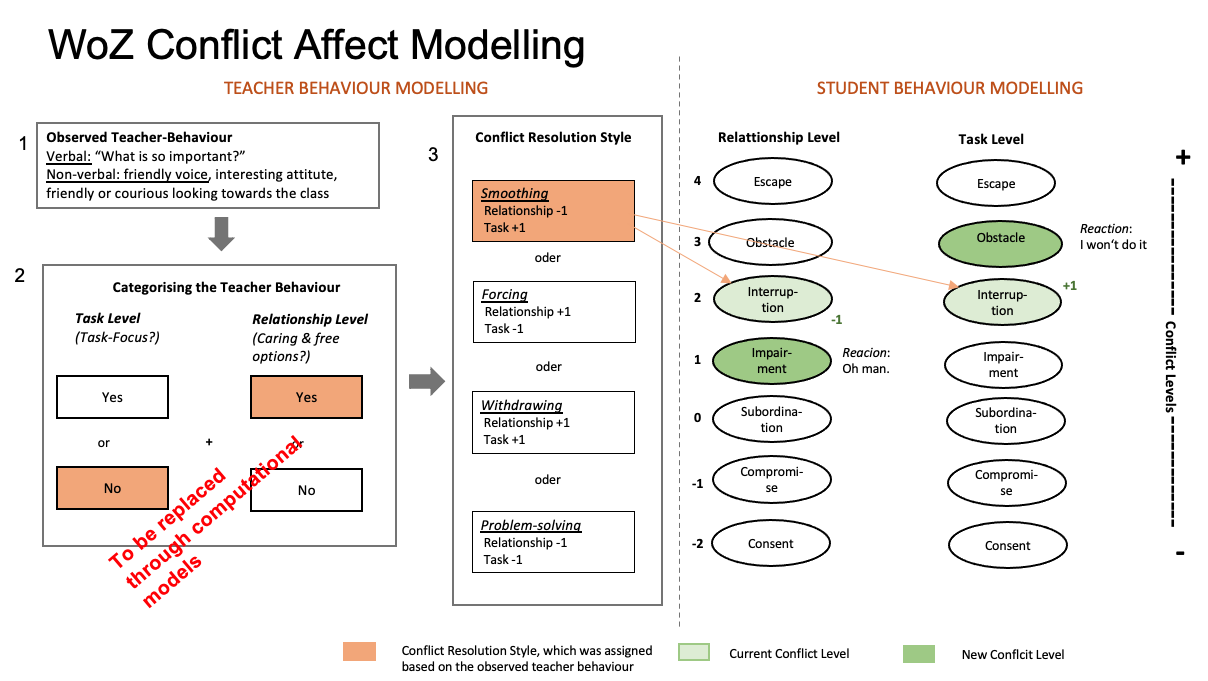}
\caption{This figure shows the mapping of the conflict resolution styles on to the different possible phases in the MITHOS Training Stage 1: First situative training.}
\label{fig_conflict_resolution}
\end{figure*}

This stage provides an interactive virtual platform for the teacher to experience dealing with complex socio-emotional challenges, such as those that can be triggered by heterogeneous learning groups or behavioral conflicts. During this stage, the teacher holds a class in front of a virtual classroom where a conflict is initiated through the behavior of one virtual student. The interaction is structured in four conflict phases to account for different phases of conflict resolution \cite{keller2001konfliktmanagement, darnon2006mastery}, Figure \ref{fig_conflict_resolution}. The virtual student’s behavior during the whole interaction with the teacher is modeled, based on a turn-by-turn evaluation of the teacher behavior on two dimensions, task and relationship conflict level \cite{darnon2006mastery}, and a matrix of conflict regulation potential ranging from -2 (low conflict potential) to +2 (high conflict potential) \cite{holt2005culture}\ref{fig_teacher_behavior_model}. The student reaction is based on the assessed conflict resolution style of the teacher. The conflict resolutions style comprises two levels: Task and Relationship \cite{keller2001konfliktmanagement}. The model  penalise (-1) or reward (+1) the teacher behaviour at both levels locally for each turn, and a global conflict level status is updated accordingly. The conflict level status is based on an ordered list which represent the conflict escalation potential of the conflict resolution style and connect the teacher behaviour to the student reaction. The concrete definitions of the levels of conflict resolutions style are based on the OPD model and represent a type of internal conflict commonly observed in situations of diversity discrimination \cite{opd2001operationalized}, and hence very likely to surface in relations of power in heterogenous classes: the control-submissiveness conflict. Therefore, at the task level, the teacher is required to involve the student in the lesson and guide them to pay attention, in a professional and respectful way. That is, the teacher should be able to maintain a balance between involving the student to a certain extent that does not violate their sense of control. At the same time, and crucial for the internal conflict, relationship level, the teacher emotionally supports the student while respecting the student's personal space and limits. 
The virtual student reactions are implemented reciprocally and are multimodal, to reflect different communication channels that include verbal and non-verbal, as well as explicit and implicit signals in the interaction. The conflict behavior of the teacher influences the conflict potential in the interaction. The teacher may successfully resolve the conflict in the final solution phase, or fail to resolve it. 

\subsection{Training stage 2 $<$\textbf{Avatar-replay stage}$>$} In this stage an avatar plays back the verbal and non-verbal behavior of the teacher allowing them to observe their reactions and take the physical perspective of the student. The avatar-replay is considered as an awareness tool, as the teacher learns about their own behavior, such as facial expressions, from the others’ perception. In this stage, the teacher is asked to watch the avatar-replay on a Monitor. The visual similarity between the teacher and avatar has been shown to play an important role in increasing self-awareness \cite{hooi2012being}. Our empirical research indicates that generic avatars may not adequately represent the self when perceptions are involved \cite{alves2023visual}. Hence, to further validate the importance of having a similar avatar and also assess the influence of behavioral contingency, we consider two groups. In the first group, the teacher watches the replay of their behavior using a default avatar. In the second group, the teacher watches the replay of their behavior on an adjusted avatar based on three images received prior to the experiment day.

\subsection{Training stage 3 $<$\textbf{Agent-feedback stage}$>$} This stage involves a virtual training agent supporting the teacher to emotionally assess their experience during the interaction with the virtual student. The stage starts with an introduction to establish trust between the teacher and the training agent \cite{10.1145/3477322}. The training agent then retrieves fragments of the interaction video where unproductive negative affections/emotionally demanding situations are detected to allow the teacher to recall the emotions felt during their experience \cite{barsalou1999perceptual}. As the teacher reflects on their experience the training agent supports the teacher in a process of co-regulation. Co-regulation in this context is the training agent’s ability to validate and understand the emotions that the teacher experienced and provide implicit or explicit empathic responses in the aim of supporting the teacher’s self-regulation \cite{murray2017promoting}. We consider the empathic response to encompass both, affective empathy which is reflecting the ability to validate and share the feelings of the teacher  and cognitive empathy or perspective taking which is simply taking the view of the teacher into consideration. We further integrate compassion within the agent’s empathic response to train the teacher on self-compassion. Self-compassion is a socio-emotional skill enabling one to direct care to oneself by engaging with one's own difficult emotions in a kind way to achieve emotional balance \cite{gilbert2014origins, neff2003self}. Self-compassion has been positively correlated with self-regulation, prosocial behavior and well-being \cite{sirois2015self}. By training teachers to be self-compassionate and supporting their ability to self-regulate during conflict situations, we increase their ability to focus on student’s emotions and learning performance. 


\begin{figure}[h]
    \centering
    
    \begin{subfigure}[b]{\columnwidth}
        \centering
        \includegraphics[width=\textwidth]{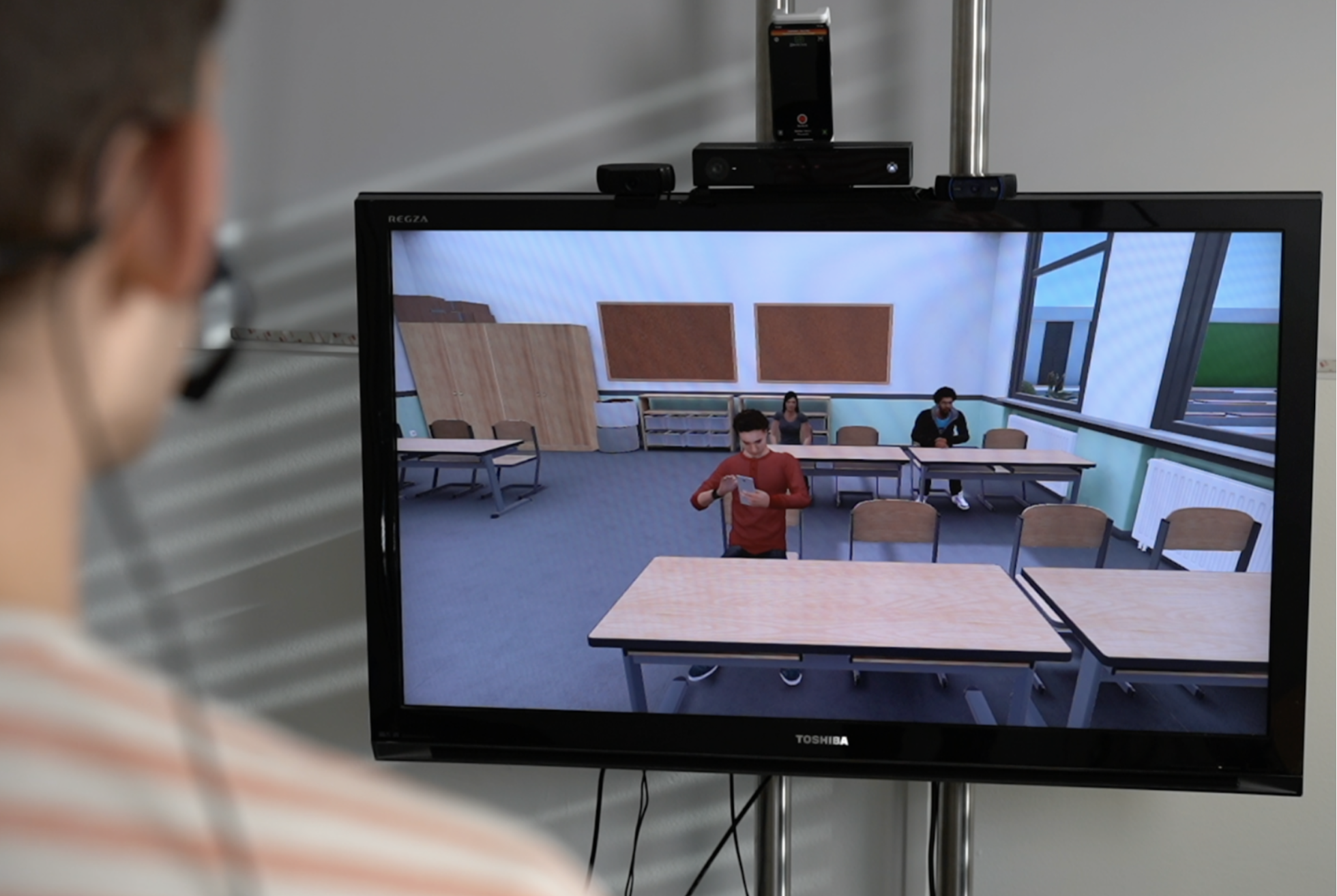}
        \caption{Participant's room}
        \label{fig_interaction_room}
    \end{subfigure}
    \hspace{0.05\columnwidth} 
    
    \begin{subfigure}[b]{\columnwidth}
        \centering
        \includegraphics[width=\textwidth]{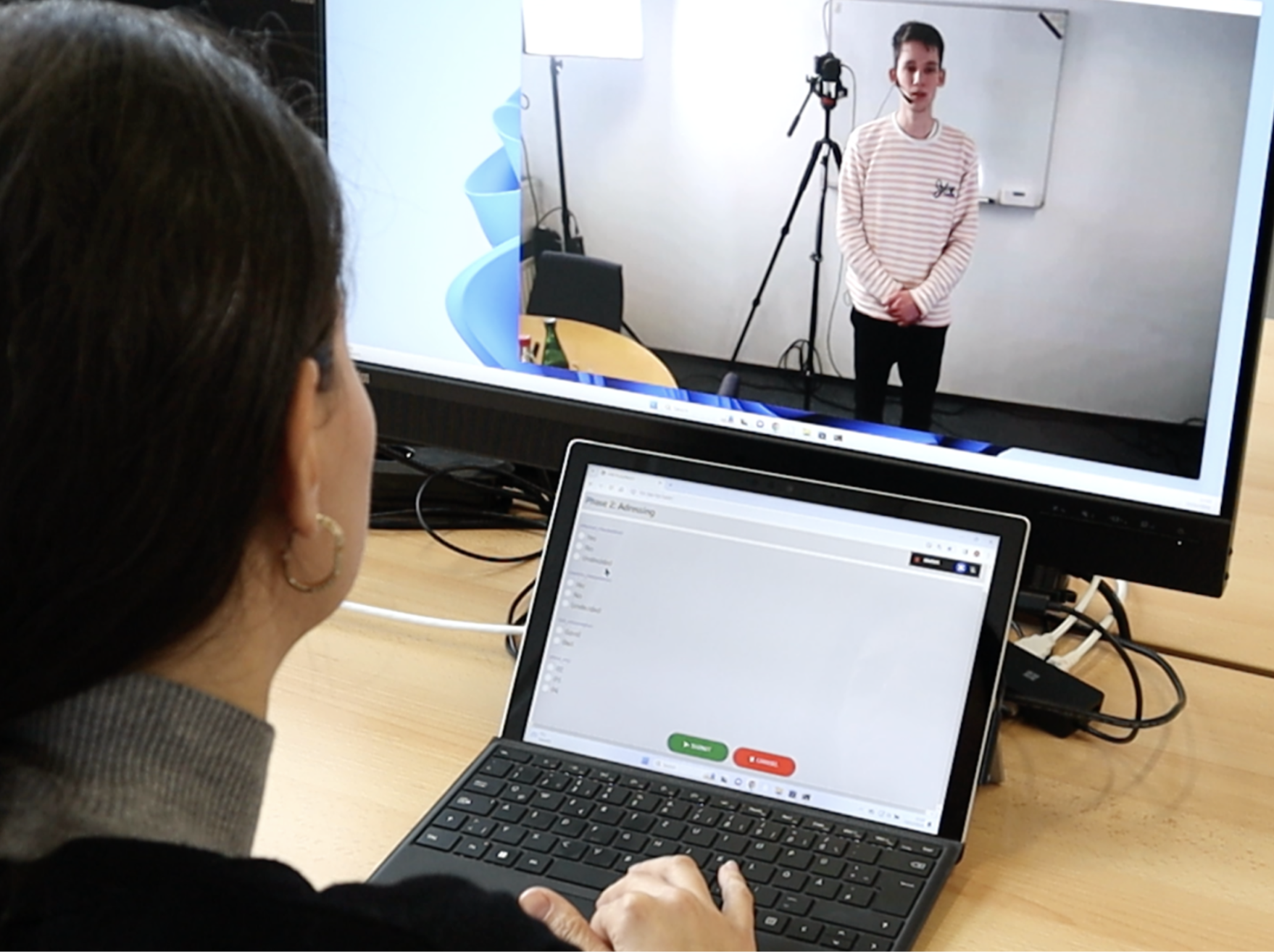}
        \caption{Wizard's room}
        \label{fig_wizard_room}
    \end{subfigure}
    
    \caption{The room setup for the Mithos Wizard of Oz (WoZ) system which is spread over two rooms.}
    \label{fig_WoZ_room_setup}
\end{figure}



\begin{figure}[!t]
\centering
\includegraphics[width=2.5in]{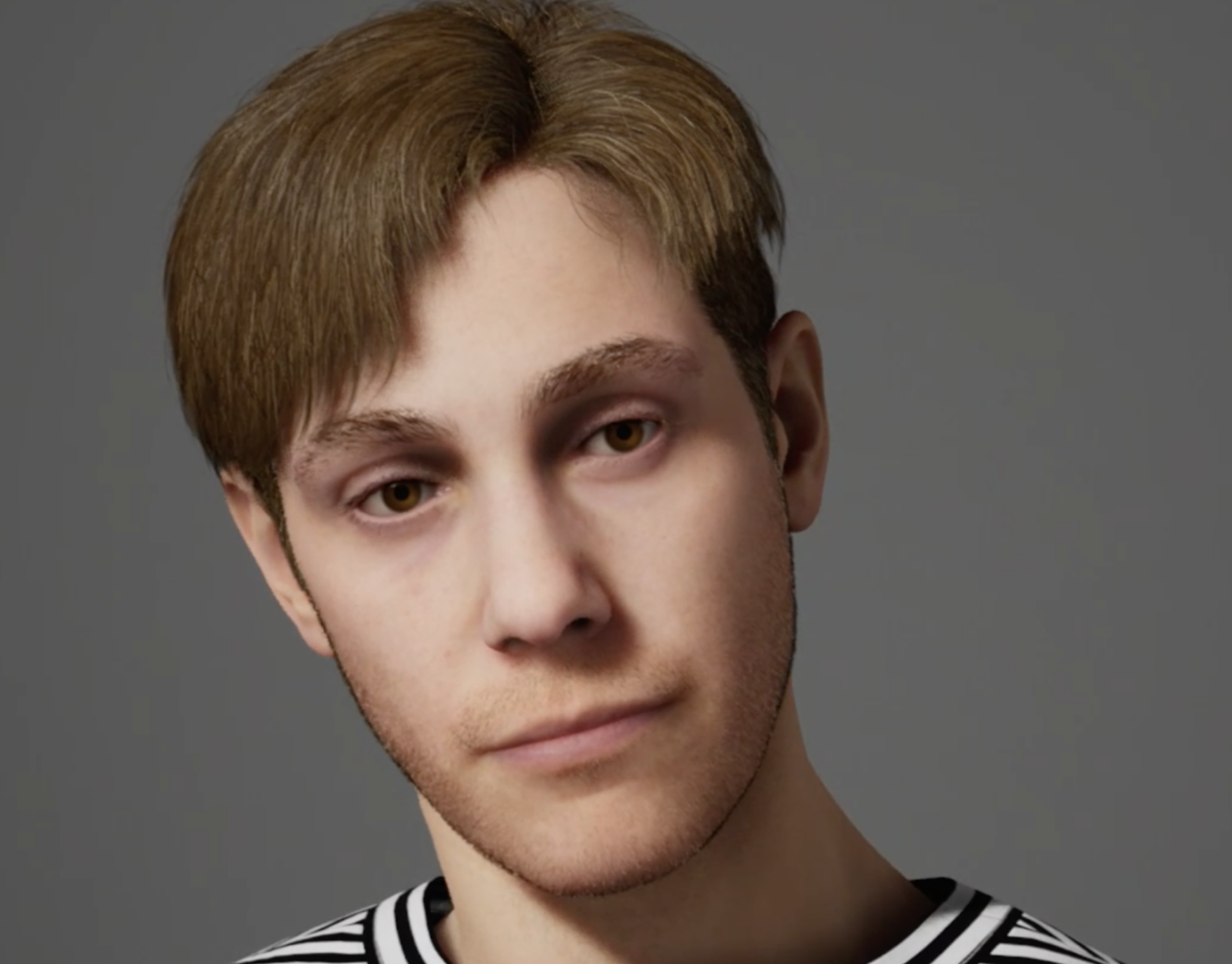}
\caption{MetaHuman virtual avatar used for Training stage 2 $<$Avatar-replay stage$>$. Note that this avatar is personalised to match each teacher participant and in this case it matches the participant in Figure \ref{fig_WoZ_room_setup}.} 
\label{fig_avatar_replay}
\end{figure}

\subsection{Training stage 4 $<$\textbf{Expert-feedback stage}$>$} 
Finally, in the Expert-feedback stage, the teacher gets verbal feedback from Human experts on positive regard as an important aspect of professional conflict regulation in the classroom. Expert feedback plays an important role for the development of professional behavior \cite{prilop2019digital, prilop2020effects, scheeler2004providing}, which also corresponds to Ericsson's theory of deliberate practice \cite{ericsson2003development}. In MITHOS, we support experts in giving feedback with a rating inventory regarding positive regard (adapted from  \cite{protzel2015wertschatzung}, and with aggregated information from the social signal analysis regarding non- or paraverbal aspects of empathy and positive regard (like eye-contact with the virtual students). These should support the expert, who observes the behavior of the student teacher in the virtual classroom, to give informed, and valuable feedback. We decided to add feedback from a human expert, due to the fact that the assessment of empathy or positive regard is actually too complex for a fully automated system.

\section{Developing Computational models for automating the system - (used in both WoZ and Automated systems)}
\subsection{Conceptual explanation of conflict model}
The virtual student sequence behavior has two end alternatives: escalation (the student gets up and leaves the classroom) and resolution (the student puts the cell phone away and turns his attention to the lesson). This is dependent on the internal conflict initially assigned to the virtual student and the behavior of the teacher. Based on the conflict regulation strategies, the teacher may attempt to resolve the situation in 5 different ways: withdraw, force, problem solve, smooth and compromise \cite{rahim2001managing}. Although compromising has been shown to be effective in conflict regulation, it stems from comparing one's interest against the other rather than considering both concerns equally.  As such, for a classroom setting and considering the well-being of both the student and the teacher, compromising is not considered as an effective regulation strategy and rather integrated in the four other strategies. 

The student's reaction is always operationalized as a shift between the conflict levels by one unit as indicated in Fig \ref{fig_conflict_resolution}. The top level represents the most suboptimal level and the lowest level represents the optimal level, where the student conforms to the request and attends to the classroom. For example, if the teacher displays a behavior that has been assigned to the conflict regulation style "Smoothing" (Task Focus = No; Attachment \& Freedom (Relationship) = Yes), the current conflict level moves up one level on the task level (towards "Flight") and down one level on the relationship level (towards “Consent”). This shift in the current conflict level is then visible in the student's reaction. A stronger refusal to cooperate in the classroom (task level) is shown, than in the previous turn but an emotionally agreeable approach to the teacher compared to the previous turn (relationship level).

\subsection{Emotional state model, Social obligations, intentions and internal conflict}
In the WoZ study, a trained human expert, the wizard, watches the interaction on a live stream and rates the participant’s performance on relationship level, task level and conflict phase. In the final VR system, the process is to be automated based on computational models developed from the gathered data during the WoZ study. 
The emotion recognition module in the Automated System utilizes what the participant decides to reveal and utilizes internal models such as \cite{gebhard2005alma} 
to understand internal emotional experiences. This module is designed based on psychological models and theories, which are validated by self-reports and qualitative analysis of the recorded data in the WoZ system for final fine-tuning in the final system. 

The self-reported data from the WoZ study, although provided directly after the experience, provides room for self-reflection and can be strongly influenced by various biases \cite{choi2023logs}. Therefore, to better understand the emotional experiences of the participants and to improve the computational models for an automated system, we conducted an interview with each participant after the experiment. The questions guides the participant to report rather personal and sensitive feelings they experienced during the interaction specifying what influences their behavior such as previous experiences and social norms. To support the participant to recall their interaction, they watched their interaction on a video. The video stopped at certain points that reflected expressions worthy of reflection. The interviewer was trained to provide a safe and trusting environment and to encourage the participant to reflect on their feelings. The interviewer validated their understanding by repeating what the participant expressed. Furthermore, self-reported data on emotions was gathered at each point during the interview as this data is rather more specific and personal and could be sensitive to talk about. The self-reported data included questions on the emotional state model \cite{mehrabian1974basic} and Lead affect (adapted from \cite{opd2001operationalized}). Qualitative analysis of the interview allowed us to assess the participant’s reaction to gather the social obligations, intentions and the internal conflict of the teacher.  

\section{Technical Implementation of MITHOS Training System For Teachers (DFKI)}
\subsection{MITHOS Wizard of Oz (WoZ) System (Visual SceneMaker - StudyMaster)}

\begin{figure*}[htbp]
    \centering
    \includegraphics[width=1\linewidth]{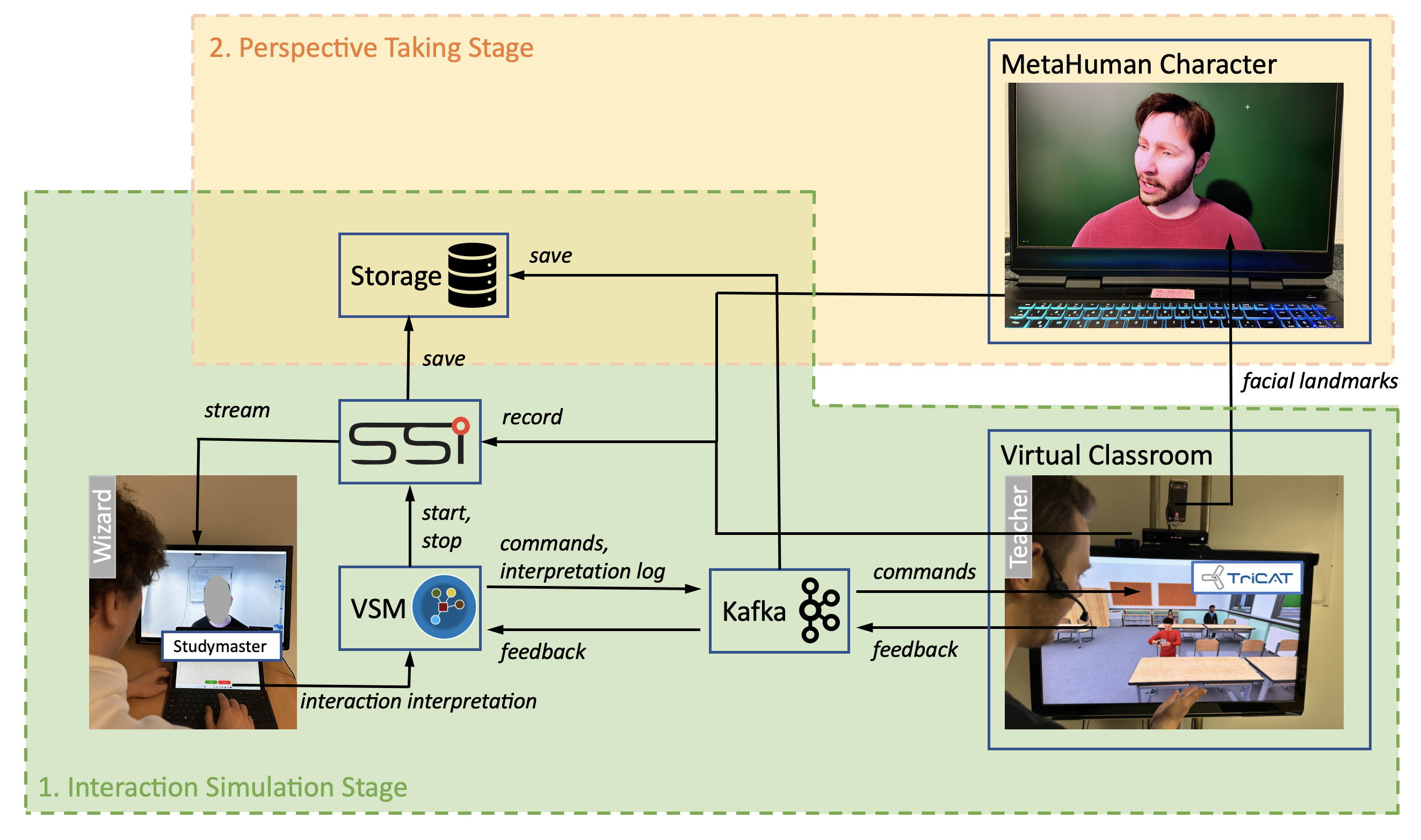}
    \caption{System architecture overview of MITHOS Wizard of Oz (WoZ) setup that depicts the setup spread over two rooms and the different system components that the participant and wizard, respectively interact with. }
    \label{fig_Woz_sys_overview}
\end{figure*}

The MITHOS WoZ system consists of hardware and software components (Fig. \ref{fig_Woz_sys_overview}) that allow the teacher to interact with the SIA, while unknowingly being monitored by the wizard. The wizard is interpreting their actions "behind the scenes". Based on this interpretation, VisualSceneMaker (VSM, see below) automatically updates the conflict level for the teacher behavior, and the conflict level and phase of the student reaction, and sends corresponding, pre-defined animation commands to the social agents in the virtual classroom for rendering.

The hardware components used are 3 computers, an iPhone, Xbox Kinect, web camera, a microphone and a heart rate sensor. AS seen in the interaction room (Figure \ref{fig_interaction_room}), one computer is used to run the Xbox Kinect, web camera, microphone and heart rate sensor. The data from these sensors are stored, as well as streamed in real-time to the Wizard in the other room (Figure \ref{fig_wizard_room}) as indicated in the Figure \ref{fig_Woz_sys_overview}. Simultaneously, as in Figure  \ref{fig_Woz_sys_overview} another computer is connected to the iPhone, which in turn detects the facial expressions of the participant during the interaction, that will be played back in the Replay Stage. At the same time, a third computer is used by the Wizard in another room (Figure  \ref{fig_wizard_room}) to evaluate the behaviour, in real-time, of the participant who is interacting with the SIA student character.

In addition to enabling real-time and natural student-teacher interaction based on expert knowledge, this setup also allows for the recording of audio, visual, postural and heart rate features that are used to build a machine learning dataset. This data is being used to develop multimodal non-verbal and verbal affect analysis machine learning systems that are used to drive computational affect models for automatically determining the current emotional experience of the teacher and to generate an appropriate student behavior. Additionally, facial data captured during the interaction is used to replay the teacher's behavior for Avatar-replay stage in the current WoZ system. 

The software components that drive this WoZ implementation are:

    \subsubsection{Apache Kafka} is an intermediary system that is used for logging participant metadata and information about the sequence of events occurring within the system, during the interaction with the participant. And this platform enables each module within the MITHOS system to communicate with each other in real-time.

    \subsubsection{VisualSceneMaker (VSM) \cite{gebhard2012visual}} VSM is a scene authoring tool which maps the wizard’s evaluation of the teacher behavior to an appropriate student behavior. The wizard’s evaluation and the different student behaviors are set up as high level abstractions and within each, there are several computations occurring to keep a track of the wizard’s evaluation of the teacher and accordingly, the exact sequences of animation commands to be executed in the 3D virtual classroom to achieve realistic interaction.
    
    \subsubsection{StudyMaster} A tool originally used for data collection \cite{schneeberger2019can, schneeberger2019would}, is now extended as a tool for orchestrating WoZ experiments \cite{bhuvaneshwara2023mithos} and has been re-implemented as a web service for remote control of agents over the internet. This provides an easy-to-use GUI for the wizard to start and stop the experiment and simultaneously start synchronized recordings of the session. However, the primary use is to send real time evaluations (see Figure \ref{fig_wizard_room}) of the teacher behavior to VSM, so that the evaluations can be mapped to student behaviors that can be animated appropriately in the virtual classroom.
    
    \subsubsection{TriCAT spaces® classroom environment} 
\begin{figure}
    \centering
    \includegraphics[width=1\linewidth]{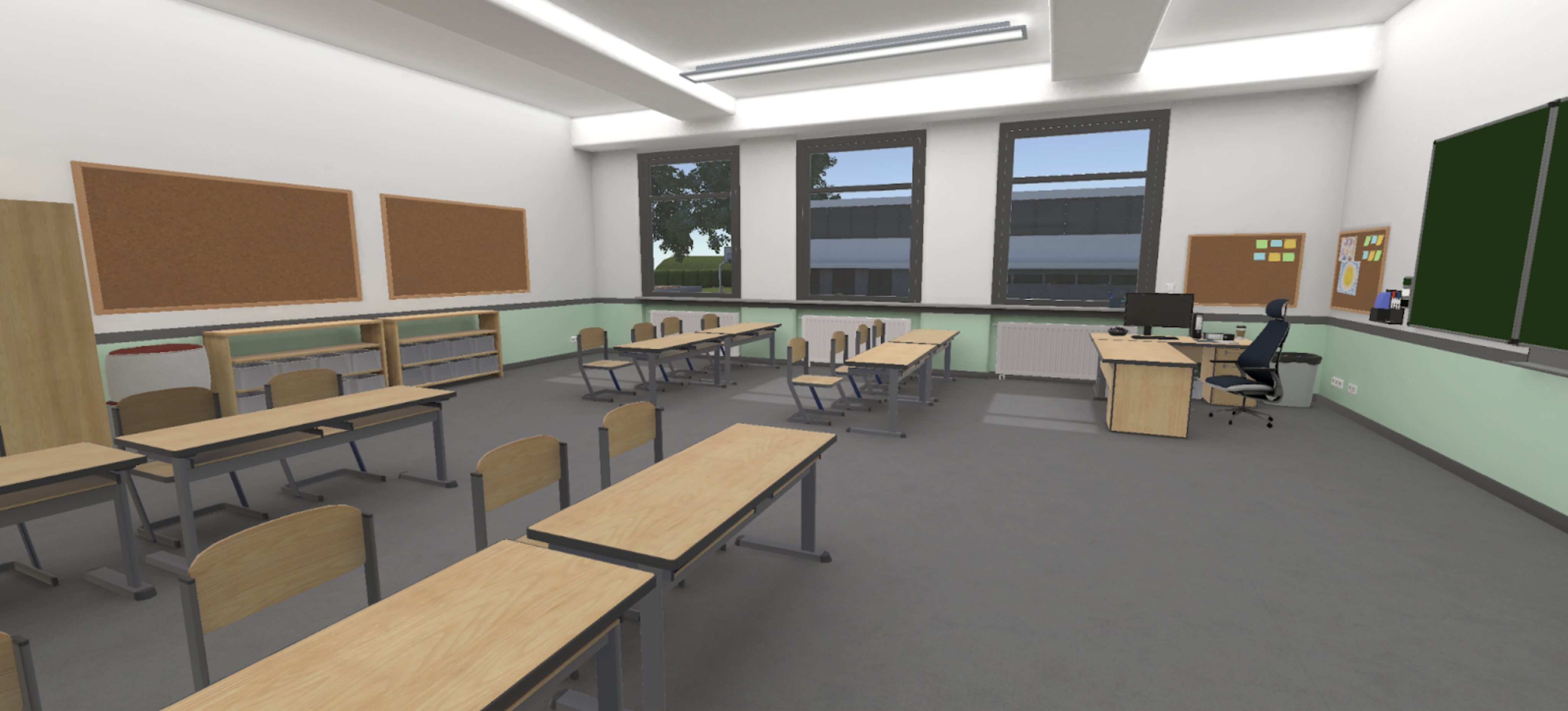}
    \caption{A snapshot from the virtual classroom setup for the MITHOS systems using TriCAT spaces®.}
    \label{fig_tricat_spaces}
\end{figure}

(Fig. \ref{fig_tricat_spaces}) is the central, virtual-immersive, multiplayer 3D-environment which can be visited by the trainees using a Desktop application as well as standard XR-HMDs. The modeled 3D-environment consists of a classroom in which the training scenario takes place. In this environment, there are three controllable student avatars, capable of performing actions such as facial expressions, walking to an object and producing speech. Although the classroom environemnt supports XR-HMDs, the WoZ study is performed in desktop mode, to ensure that we collect all potentially relevant facial and postural features in our MITHOS dataset. This is done deliberately to ensure that we can make better decisions for the final automated system by observing the participants without any items restricting their facial view.

    During the training scenarios, the classroom is populated by several interactive student agents whose multimodal behavior is controlled by executing specially designed scenario activity scripts. These scripts specify the verbal utterances of the agents together with coverbally aligned, speech-accompanying non-verbal behavior, such as facial expressions, gestures, poses or other actions in the virtual environment, like walking to an object or another agent, sitting down on a chair, interacting with a smartphone, and many more. 
    


    \subsubsection{Social Signal Interpretation (SSI)\cite{10.1145/2502081.2502223}} framework is used to store data streams of applied sensor devices during recording sessions. Its strength lies in the guaranteed synchronization in time across several sensors and even locally separated devices. SSI is also used for real-time streaming of audiovisual data between networked computer devices. It enables the wizard to observe the virtual student and human teacher interaction in real time. This is required as the MITHOS WoZ system is spread over two different rooms: One room for student-teacher interaction system (Fig. \ref{fig_interaction_room}) and another room with the wizard (Fig. \ref{fig_wizard_room}) observing the live feed of the student-teacher interaction from the former room.
    
    \subsubsection{MetaHuman, Unreal Engine, Qixel Bridge and Live Link Face} to map the facial expressions of the teacher onto a MetaHuman virtual avatar for Training stage 2 $<$Avatar-replay stage$>$ (Fig. \ref{fig_avatar_replay}). This is done by recording the behavior of the teacher during the interaction using Live Link Face app on an iPhone and driving the facial expressions of a connected MetaHuman character in Unreal Engine. Qixel Bridge is used to download, to Unreal Engine, the MetaHuman character which can possibly be designed to look like the current participant teacher.

\subsection{MITHOS Automated Training System}
The MITHOS Automated Training System is developed using the data collected from the study conducted using the MITHOS WoZ system. Besides the components in the WoZ data collection system, the final automated MITHOS training system, shown in Figure \ref{fig_automated_sys_overview} is set up in a single room with the HTC Vive and associated trackers, that are seen in Figure \ref{fig_vive_HMD}, all connected to the one computer. There are several software components that are required to enable the full functioning of this automated system and these components are as follows: 


\begin{figure*}[htbp]
    \centering
    \includegraphics[width=\textwidth]{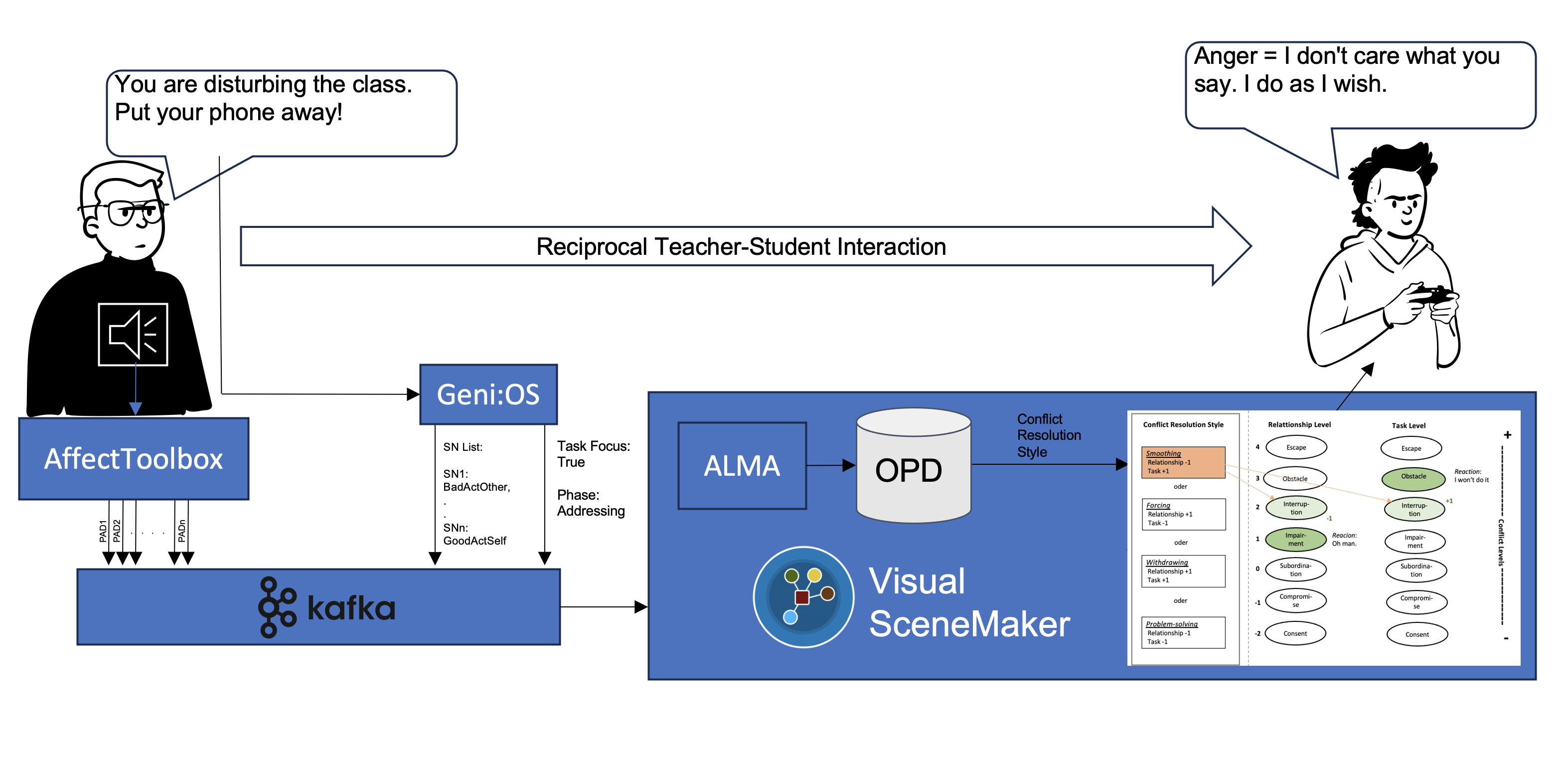}
    \caption{System architecture overview of MITHOS Automated Setup.The figure shows that the voice and language content of the teacher are processed by AffectToolbox and Geni:OS respectively. The results are passed to Visual Scenemaker that uses the ALMA and OPD models to obtain the final conflict regulation style and drive the behaviour of the SIA student automatically.}
    \label{fig_automated_sys_overview}
\end{figure*}

    \subsubsection{Apache Kafka} is continued to be used as the intermediary system for logging participant metadata and information about the sequence of events occurring within the modules of the automated system, during the interaction with the participant.

    \subsubsection{TriCAT spaces® VR enabled classroom environment} is the central, virtual-immersive, 3D-environment which can be visited by the participating teachers in Virtual Reality mode with the HTC Vive headset, through a desktop application. 
    
\begin{figure}[!t]
    \centering
    \includegraphics[width=2.5in]{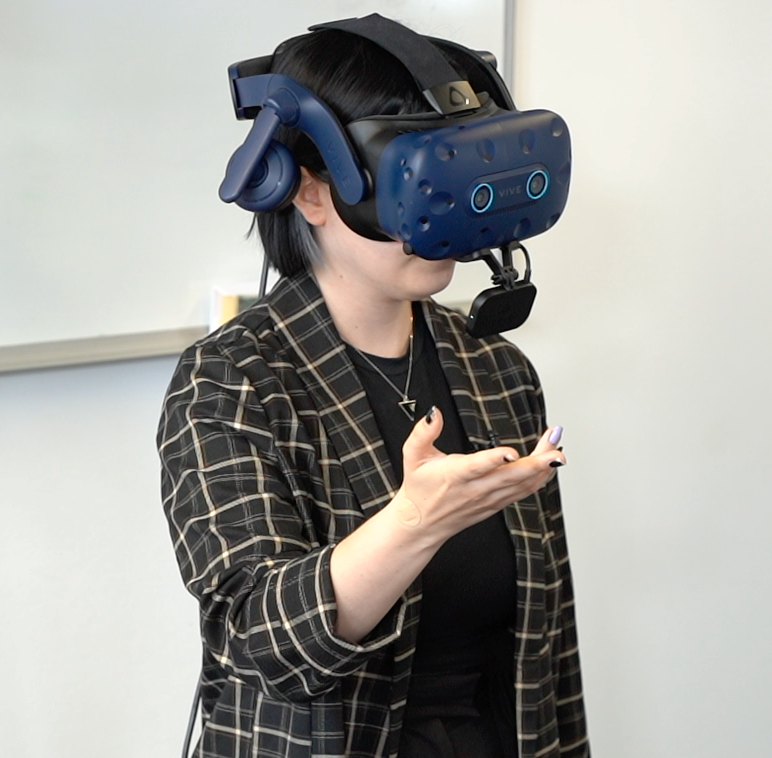}
    \caption{This figure shows a participant teacher interacting with the VR classroom with the help of the HTC Vive Pro Eye Headset that is connected to HTC Vive Facial Tracker.}
    \label{fig_vive_HMD}
    \end{figure}

    
    The system underlying the virtual-immersive environment is also able to detect, pre-process and interpret a variety of user inputs and interaction events in the environment and forward them to the corresponding topics maintained in Apache Kafka service. A voice activity detection and an automatic speech recognition service are responsible for transcribing the users’ verbal utterances into the corresponding textual passages. The transcriptions as well as the resulting interpretation hypotheses are then forwarded to the Apache Kafka server. An interpersonal distance recognition component detects the users’ proxemic distance to the agents or other users in the scene and generates events representing different proxemics categories, such as an intimate, personal, social and public zone. Furthermore, the system is able to process captured data from a gaze tracker, a facial expression tracker (shown in Figure \ref{fig_vive_HMD}) as well as several body trackers. 
    
   In the Interaction Simulation Stage, the pupil dilation, blinking intervals and eye movements are pre-processed and recorded. Similarly, the facial expression tracker also captures the user’s facial movements and this is also stored along with the participant's voice. In the Avatar-replay Stage, the recorded information can be played back from the perspective of the student, so that the participating teacher can observe their own behaviour, with recreated eye movement and facial expressions. This entire VR setup is built using the HTC VIVE stack shown in Figure \ref{fig_vive_HMD}.

    \subsubsection{AffectToolbox \cite{mertes2024affecttoolbox}} is used in the context of MITHOS to generate the observable emotion understanding information of the teacher behaviour in the Pleasure, Arousal and Dominance (PAD) dimensional space \cite{mehrabian1974basic}. The AffectToolbox is essentially an open source system that is used for real-time recognition of a user’s affective state. It is able to generate multiple multimodal data streams (i.e. audio, transcript, video and skeleton data) from audiovisual sensory devices (e.g. webcams). Respective modalities are then analyzed with integrated machine learning techniques. The uni-modal results of applied affect recognition models are represented by a subset of pleasure, arousal and/or dominance scores. These unimodal emotional cues are the input for an event-driven fusion algorithm, which deduces a coherent emotional state, represented in the continuous PAD emotional space, which is written to a relevant Kafka topic and will be used to further comprehend the affective state of the participant based on indicators in the verbal behaviour, classroom social norms and the decay of emotions with time. 
    
    \subsubsection{ Geni:OS} is a software framework for multimodal intelligent assistants which is used in MITHOS for natural language understanding, integration of input modalities, and the evaluation of social norms (SNs) relevant in the classroom. This input processing replaces the phase and task focus interpretations by the wizard in the previously described WoZ system. Incoming data streams for gaze, and interpersonal distance (from TriCAT spaces) are pre-processed and aggregated. Speech transcripts are fed into a natural language understanding module. The geni:OS multimodal fusion component integrates the information from all modalities to compound intervals called Interaction Acts that represent coherent units of the user’s communicative behavior. The following integration rules are active for the different modalities:
    
    Each spoken utterance is considered as an interpretable interval, i.e. it generates a possible Interaction Act which is filled with the result of Geni:OS’s natural language understanding component. It uses a machine learning model to analyze the result of the speech recognition tool and derive a speech act. The model was trained on teacher utterances collected from real classroom situations and from the MITHOS WoZ data collection study. 
    
    
    
    Once the multimodal fusion has created an Interaction Act representing the teacher’s last meaningful conversational unit, it gets analyzed in the context of the classroom situation. To this end, a Discourse Analysis component \cite{rubenDiscourseAnalyser} is fed with the virtual student’s behaviors as well as with the teacher’s Interaction Act. It evaluates the teacher’s behavior based on a set of social norms for classroom situations and general interaction. As a result, a set of violated and a set of adhered-to social norms is provided, which Geni:OS uses to assign an ALMA-style appraisal tag (cf. below) to the Interaction Act based. The thus enriched Interaction Act  is sent to VSM.
    
    \subsubsection{VSM and ALMA} As in the WoZ study, all the final decisions to produce the SIA student behaviour is done in the Visual SceneMaker (VSM). As seen in Figure \ref{fig_automated_sys_overview} VSM receives multiple PAD updates from the AffectToolbox and it also receives the Interaction Acts from Geni:OS. The updates from both of these components can be fed into the emotion model  ALMA \cite{gebhard2005alma} that is running within the scope of VSM. ALMA uses all the AffectToolbox updates from within one Interaction Act and the information from the Interaction Act itself to simulate the teachers regulated (i.e. perceivable to the outside) emotion in real time. The detected regulated emotion is then mapped to the concept of Lead Affect that is driving the person's behaviour. The mapping is based on the model developed using OPD \cite{opd2001operationalized} and ALMA \cite{gebhard2005alma} and is designed to produce the information about the relationship level. The information about the task level and phase of the interaction is supplied by Geni:OS. VSM puts all of this information together to produce the conflict regulation style and therefore drive the student behaviour as shown in Figure \ref{fig_automated_sys_overview}.

\section{Empirical Validations of the Interaction Scenario and Evaluation of the Avatar-User Similarity}
We conducted two evaluation studies for piloting (a) the authenticity of the interaction behavior of the virtual student and (b) the effects of similarity of the virtual agent with the participant for the avatar-replay stage.

\subsection{Study 1: Interaction Scenario}
For the effectiveness of a virtual training, the authenticity of the simulated scenario is of high importance to evoke comparable cognitive processes (like norms and expectations) and affective reactions (like stress) as an interaction with a real student in a real classroom setting will do \cite{poeschl2013german}. If a virtual scenario is seen as realistic and authentic by the users, it can result in a feeling of presence, meaning that people feel, think and behave the way they do in a similar real world situation \cite{pertaub2002experiment}. To assess the authenticity of the scenario, we created three videos of a first demonstrator with reactions of the virtual student. The videos showed different conflict solutions and consisted of 10 to 11 reactions of the virtual student. The duration of reactions varies between 7 and 15 seconds and is about 2.5 minutes for each of the three complete scenes. VSM automatically generates a combination of possible behaviors of the virtual student depending on the stage of the conflict and the rating of the Wizard. Therefore, over 500 behavior combinations could be possible. This includes combinations with similar conflict levels on the task- and relationship-dimension and combinations with different conflict levels of the two dimensions. To evaluate further, whether the difference/similarity of conflict levels has an influence on authenticity, we selected reactions for (1) a similar conflict level and (2) a high difference of conflict level between task and relationship.

\subsubsection{Participants and Design}

The N=57 participants (n=51 preservice- and n=6 in-service teachers) had the instruction to assess at first each of the reactions of the student individually, and after that the whole video scene is replayed and assessed.

The assessment of the reactions of the interacting student includes three questions of the authenticity of (1) the posture and gesture, (2) the language, and (3) the behavior (example: “The behavior of the virtual student corresponds to the behavior of a real student”). After watching the whole scene, the participants rated the authenticity of the background students with four items regarding posture and gesture, behavior, and comparability with real students (example: “The behavior of the students in the class corresponds to the behavior of real students”). Further, the whole scene should be rated with three items (Item stem: “I found the shown videos as…” “…meaningful”, “…untypical” and “…authentic”). All items were rated on a 5-point Likert scale from “not correct at all” to “fully correct”. The items were adapted from \cite{fitrianie202019, langer2022entwicklung, poeschl2013german, seidel2010observer, seufert2022classroom, slater1999measuring}, Cronbachs alpha ranged from .89 to .96.

\subsubsection{Results}

The rating of the authenticity was for all scales above the mean scale value (M=3.24, SD=.57; background students: M=3.47, SD=.64, whole scene: M=3.09, SD=.81). To evaluate, if we could find a difference of authenticity of the behavior of the interacting student depending on the similarity of the conflict level of task and relationship dimension, we performed an ANOVA with repeated measurement on the factor similarity (high vs. low). We found a statistically significant difference, meaning that the reactions with a low similarity are less authentic than reactions with high similarity of the two dimensions (High similarity: M=3.34, SD=.62; low similarity: M=3.19, SD=.56; F(1,56)=8.01, p$<$.01).

\subsubsection{Discussion}

A VR-Classroom should be authentic so that people feel, think and behave like they do in a similar situation in the real world, and therefore training can be effective. Our results showed a medium authenticity of the first scenario; further we found a difference of authenticity depending on the similarity of the conflict level between task and relationship dimension. To optimize the authenticity of the MITHOS scenario, the movements of the virtual students will be smoothened, we will implement a natural voice (see chapter 2c) and will adapt the content of the spoken sentences of the interacting student. 

\subsection{Study 2: Effects of Agent-User Similarity}
As we use an avatar (virtual agent representing the teacher-user) as an awareness tool during the avatar-replay stage to increase the teacher’s self-awareness, we investigate the effect of varying degrees of avatar-person similarity and its influence on user-identification and affinity. Using a theory-based methodology, we looked beyond the distinction between a personalized vs. default avatar and included three conditions. We assessed the perceived facial similarity between avatar and user, explicit-cognitive and implicit-emotional identification with the avatar and perceived realism of the avatar. 

\subsubsection{Participants and Design}

The study was conducted online and included 33 participants (13 males and 20 females) of age range between 18 and 40 years (M = 27.6 years, SD = 4.74 years).  The conditions ranged from the most similar having 0\% manipulation of facial features to two other conditions where manipulation increased by 50\% and 100\% and an additional control group (CG). 

Participants filled out a questionnaire to collect their email, three facial images (as required by FaceGen) and demographic data. The avatars for each condition were created in 3D and added in the form of animated images (GIF) representing idle behavior. Then, the participant received the second questionnaire where the conditions and instruments were presented to the participants in random order. The instruments included 5 items on ’physical similarity’ from the polythetic identification model \cite{downs2019polythetic}, three created questions on explicit cognitive identification (‘If I look at the GIF, it feels like I am this avatar’, ’I identify with this avatar’ and ’I feel represented by this avatar’ based on \cite{downs2019polythetic, van2010player},  affinity scale to assess the implicit emotional identification \cite{seymour2021have} and a single item ’How realistic is the avatar?’ on perceived realism for manipulation check.

\subsubsection{Results}

Our results partially confirmed the effect of manipulation of avatar-person similarity on user-identification and affinity.  

Effect of avatar-person similarity manipulation on perceived similarity revealed a significant main effect, F(1.87,59.72) = 10.13, p $<$ .001, $\eta$2 = .11 Subsequent comparisons did not show a significant effect between the 0\%- (M = 3.34, SD = 1.26) and the 50\%-condition (M = 3.18, SD = 1.21), t(32) = 1.36, p = .18. All other comparisons showed a significant difference, showing a steady decline in similarity with increasing similarity manipulation (0\% = highest - CG = lowest).

Similarly, the results revealed a significant main effect of avatar-person similarity manipulation on explicit identification, F(2.03,65.01) = 13.11, p $<$ .001, $\eta$2 = .13. Subsequent comparisons did not show a significant effect between the 0\%- (M = 2.76, SD = 1.84) and the 50\%-condititon (M = 2.64, SD = 1.72), t(32) = 0.64, p = .53, or between the 100\%- (M = 1.82, SD = 1.05) and the CG (M = 1.53, SD = 0.67), t(32) = 1.68, p = .1. All other comparisons showed a significant difference, showing a decline in explicit identification with increasing similarity manipulation.

The results also revealed a significant main effect of the degree of avatar-person similarity manipulation (condition) on affinity, F(2.2,70.44) = 7.66, p $<$ .001, $\eta$2 =.05. Subsequent comparisons again did neither find a significant effect between the 0\%- (M = 3.22, SD = 1.77) and the 50\%-group (M = 3.08, SD = 1.72), t(32) = 0.95, p = .35, nor between the 100\%- (M = 2.52, SD = 1.51) and the CG (M = 2.37, SD = 1.3), t(32) = 0.74, p = .46. However, all the other group comparisons showed a significant difference, showing a decline in explicit identification with increasing similarity manipulation.

\subsubsection{Discussion}

Similar to previous research, the results indicate that avatar-person identification corresponds with perceived visual similarity. Including the different degrees of similarity shows that generic avatars are insufficient when studying concepts related to self-perception such as self-awareness. Moreover, the insignificance between 0\% and 50\% degree of manipulation gives researchers a margin of error to work with when personalizing avatars for each participant. In MITHOS, we are creating personalized avatars as one condition in the avatar-replay stage to compare to default avatars. In this case, the avatar-replay simulates the behavior of the participant during an interaction. The results would show the influence of similarity on self-awareness within the context of behavioral contingency.

\section{Conclusion and Future Work}


The MITHOS project developed an MR-system to tackle the current lack of trainings for professional communication and conflict resolution in Teacher Education. It leverages XR technologies to offer a situated training,  while avoiding disadvantages of training with actual students in classrooms. Various technologies and modelling methods are employed to this end in an interdisciplinary endeavor which is supported by specifically designed technology.       

Automated immersive VR environments developed based on social interaction and psychological models represent a novel combination to simulate interactive virtual agents and avatars. The MITHOS project applies the technology and extends existing computational models to include training teachers on socio-emotional skills to cope with challenging classroom situations. It specifically allows the training on empathic conflict regulation and adoption of the student’s perspective. 

MITHOS aims to have fully autonomous and adaptive SIA, based on an extension of the ALMA \cite{gebhard2005alma} 
computational model of emotion, mood personality and emotion regulation. MITHOS will then be able to track the teachers' regulated and unregulated emotions, and the conflict potential during the interaction, to trigger the SIA reaction. The emotional modeling will further be used to increase implicit and explicit self-awareness by direct feedback. A geni:OS component is integrated in the system and will analyze the teacher's communicative behavior and affect display concerning speech, gaze, body movement, and interpersonal distance to provide the input to ALMA. 

MITHOS is transferable for any social conflict situation and for training other population, beyond teachers in school, but also outside school, for professional development. This is due to the modular and hybrid design of the system, which allows representing different domains of human-human interaction. It is also possible, albeit with some considerable effort, to use alternative and even additional  psychological models. 

Further evaluations are running, which are testing the effects of the overall automated feedback and evaluate the individual stages. The results currently provide evidence for the MITHOS concept of supporting perspective taking through the situative mixed-reality self-compassion training. 

\section{Acknowledgments}
This study is funded by the German Federal Ministry of Education and Research within the funding line “Interactive systems in virtual and real spaces - Innovative technologies for the digital society” (Project MITHOS, grant 16SV8687).


\onecolumn
\bibliographystyle{IEEEtran}
\bibliography{refs}

\begin{thebibliography}{100}
\providecommand{\url}[1]{#1}
\csname url@samestyle\endcsname
\providecommand{\newblock}{\relax}
\providecommand{\bibinfo}[2]{#2}
\providecommand{\BIBentrySTDinterwordspacing}{\spaceskip=0pt\relax}
\providecommand{\BIBentryALTinterwordstretchfactor}{4}
\providecommand{\BIBentryALTinterwordspacing}{\spaceskip=\fontdimen2\font plus
\BIBentryALTinterwordstretchfactor\fontdimen3\font minus \fontdimen4\font\relax}
\providecommand{\BIBforeignlanguage}[2]{{%
\expandafter\ifx\csname l@#1\endcsname\relax
\typeout{** WARNING: IEEEtran.bst: No hyphenation pattern has been}%
\typeout{** loaded for the language `#1'. Using the pattern for}%
\typeout{** the default language instead.}%
\else
\language=\csname l@#1\endcsname
\fi
#2}}
\providecommand{\BIBdecl}{\relax}
\BIBdecl

\bibitem{hargreaves2000mixed}
A.~Hargreaves, ``Mixed emotions: Teachers’ perceptions of their interactions with students,'' \emph{Teaching and teacher education}, vol.~16, no.~8, pp. 811--826, 2000.

\bibitem{nathanson2014affect}
D.~L. Nathanson, ``Affect theory and the compass of shame,'' in \emph{The widening scope of shame}.\hskip 1em plus 0.5em minus 0.4em\relax Routledge, 2014, pp. 339--354.

\bibitem{gergely1996social}
G.~Gergely and J.~S. Watson, ``The social biofeedback model of parental affect-mirroring,'' \emph{The International journal of psycho-analysis}, vol.~77, no.~6, p. 1181, 1996.

\bibitem{frenzel2021teacher}
A.~C. Frenzel, L.~Daniels, and I.~Buri{\'c}, ``Teacher emotions in the classroom and their implications for students,'' \emph{Educational Psychologist}, vol.~56, no.~4, pp. 250--264, 2021.

\bibitem{gross2014emotion}
J.~J. Gross \emph{et~al.}, ``Emotion regulation: Conceptual and empirical foundations,'' \emph{Handbook of emotion regulation}, vol.~2, pp. 3--20, 2014.

\bibitem{Dale2015}
D.~Dale and C.~James, ``The importance of affective containment during unwelcome educational change: The curious incident of the deer hut fire,'' \emph{Educational Management Administration and Leadership}, vol.~43, pp. 92--106, 1 2015.

\bibitem{eckstein2015adventure}
F.~Eckstein and U.~R{\"u}th, ``Adventure-based experiential therapy with inpatients in child and adolescent psychiatry: An approach to practicability and evaluation,'' \emph{Journal of Adventure Education \& Outdoor Learning}, vol.~15, no.~1, pp. 53--63, 2015.

\bibitem{keller2001konfliktmanagement}
G.~Keller, ``Konfliktmanagement in der schule: moderieren, l{\"o}sen, vorbeugen,'' \emph{(No Title)}, 2001.

\bibitem{park2022implicit}
S.~Park and D.~Tsovaltzi, ``Implicit and explicit emotion regulation for conflict resolution: Narrative and self-compassion as anti-bullying training,'' in \emph{Proceedings of the 16th International Conference of the Learning Sciences-ICLS 2022, pp. 187-194}.\hskip 1em plus 0.5em minus 0.4em\relax International Society of the Learning Sciences, 2022.

\bibitem{krause2013arbeitssituation}
A.~Krause, C.~Dorsemagen, and S.~Baeriswyl, \emph{Zur Arbeitssituation von Lehrerinnen und Lehrern: Ein Einstieg in die Lehrerbelastungsund-gesundheitsforschung}.\hskip 1em plus 0.5em minus 0.4em\relax Springer, 2013.

\bibitem{10.1145/3477322}
B.~Lugrin, C.~Pelachaud, and D.~Traum, \emph{The Handbook on Socially Interactive Agents: 20 years of Research on Embodied Conversational Agents, Intelligent Virtual Agents, and Social Robotics Volume 1: Methods, Behavior, Cognition}, 1st~ed.\hskip 1em plus 0.5em minus 0.4em\relax Association for Computing Machinery, 2021, vol.~37.

\bibitem{fonagym}
M.~G. Fonagy, M.~Jurist, and M.~Target, ``M.(2002) affect regulation, mentalization, and the development of the self.''

\bibitem{hartmann2023imagine}
C.~Hartmann, Y.~Orli-Idrissi, L.~C.~J. Pflieger, and M.~Bannert, ``Imagine \& immerse yourself: Does visuospatial imagery moderate learning in virtual reality?'' \emph{Computers \& Education}, vol. 207, p. 104909, 2023.

\bibitem{gyurak2011explicit}
A.~Gyurak, J.~J. Gross, and A.~Etkin, ``Explicit and implicit emotion regulation: A dual-process framework,'' \emph{Cognition and emotion}, vol.~25, no.~3, pp. 400--412, 2011.

\bibitem{lenske2015linzer}
G.~Lenske and J.~Mayr, ``Das linzer konzept der klassenf{\"u}hrung (lkk). grundlagen, prinzipien und umsetzung in der lehrerbildung,'' \emph{Jahrbuch f{\"u}r Allgemeine Didaktik}, vol. 2015, pp. 71--84, 2015.

\bibitem{kessler2010embodied}
K.~Kessler and L.~A. Thomson, ``The embodied nature of spatial perspective taking: Embodied transformation versus sensorimotor interference,'' \emph{Cognition}, vol. 114, no.~1, pp. 72--88, 2010.

\bibitem{fonagyaffect}
P.~Fonagy, G.~Gergely, E.~L. Jurist, and M.~Targe, ``Affect regulation, mentalization, and the development of the self (book review),'' 2004.

\bibitem{batson1997perspective}
C.~D. Batson, S.~Early, and G.~Salvarani, ``Perspective taking: Imagining how another feels versus imaging how you would feel,'' \emph{Personality and social psychology bulletin}, vol.~23, no.~7, pp. 751--758, 1997.

\bibitem{hooi2014avatar}
R.~Hooi and H.~Cho, ``Avatar-driven self-disclosure: The virtual me is the actual me,'' \emph{Computers in Human Behavior}, vol.~39, pp. 20--28, 2014.

\bibitem{moser1996entwicklung}
U.~Moser and I.~Von~Zeppelin, ``Die entwicklung des affektsystems,'' \emph{Psyche}, vol.~50, no.~1, pp. 32--84, 1996.

\bibitem{gebhard2005alma}
P.~Gebhard, ``Alma: a layered model of affect,'' in \emph{Proceedings of the fourth international joint conference on Autonomous agents and multiagent systems}, 2005, pp. 29--36.

\bibitem{opd2001operationalized}
O.~W. Group and A.~zur Operationalisierung Psychodynamischer~Diagnostik, \emph{Operationalized Psychodynamic diagnostics: foundations and manual}.\hskip 1em plus 0.5em minus 0.4em\relax Seattle; Toronto: Hogrefe \& Huber, 2001.

\bibitem{pilling2020long}
S.~Pilling, P.~Fonagy, E.~Allison, P.~Barnett, C.~Campbell, M.~Constantinou, T.~Gardner, N.~Lorenzini, H.~Matthews, A.~Ryan \emph{et~al.}, ``Long-term outcomes of psychological interventions on children and young people’s mental health: A systematic review and meta-analysis,'' \emph{PloS one}, vol.~15, no.~11, p. e0236525, 2020.

\bibitem{greenwald2017technology}
S.~W. Greenwald, A.~Kulik, A.~Kunert, S.~Beck, B.~Fr{\"o}hlich, S.~Cobb, S.~Parsons, N.~Newbutt, C.~Gouveia, C.~Cook \emph{et~al.}, ``Technology and applications for collaborative learning in virtual reality.''\hskip 1em plus 0.5em minus 0.4em\relax Philadelphia, PA: International Society of the Learning Sciences., 2017.

\bibitem{lerner2020immersive}
D.~Lerner, S.~Mohr, J.~Schild, M.~G{\"o}ring, T.~Luiz \emph{et~al.}, ``An immersive multi-user virtual reality for emergency simulation training: Usability study,'' \emph{JMIR serious games}, vol.~8, no.~3, p. e18822, 2020.

\bibitem{dalinger2020mixed}
T.~Dalinger, K.~B. Thomas, S.~Stansberry, and Y.~Xiu, ``A mixed reality simulation offers strategic practice for pre-service teachers,'' \emph{Computers \& Education}, vol. 144, p. 103696, 2020.

\bibitem{batrinca}
L.~Batrinca, G.~Stratou, A.~Shapiro, L.-P. Morency, and S.~Scherer, ``Cicero - towards a multimodal virtual audience platform for public speaking training,'' in \emph{Intelligent Virtual Agents}, R.~Aylett, B.~Krenn, C.~Pelachaud, and H.~Shimodaira, Eds.\hskip 1em plus 0.5em minus 0.4em\relax Berlin, Heidelberg: Springer Berlin Heidelberg, 2013, pp. 116--128.

\bibitem{bautista2015exploring}
N.~U. Bautista and W.~J. Boone, ``Exploring the impact of teachme™ lab virtual classroom teaching simulation on early childhood education majors’ self-efficacy beliefs,'' \emph{Journal of Science Teacher Education}, vol.~26, no.~3, pp. 237--262, 2015.

\bibitem{tsovaltzi2014group}
D.~Tsovaltzi, T.~Puhl, R.~Judele, and A.~Weinberger, ``Group awareness support and argumentation scripts for individual preparation of arguments in facebook,'' \emph{Computers \& Education}, vol.~76, pp. 108--118, 2014.

\bibitem{puhl2015long}
T.~Puhl, D.~Tsovaltzi, and A.~Weinberger, ``A long-term view on learning to argue in facebook: The effects of group awareness tools and argumentation scripts.''\hskip 1em plus 0.5em minus 0.4em\relax International Society of the Learning Sciences, Inc.[ISLS]., 2015.

\bibitem{tsovaltzi2017leveraging}
D.~Tsovaltzi, R.~Judele, T.~Puhl, and A.~Weinberger, ``Leveraging social networking sites for knowledge co-construction: Positive effects of argumentation structure, but premature knowledge consolidation after individual preparation,'' \emph{Learning and Instruction}, vol.~52, pp. 161--179, 2017.

\bibitem{baumert2011kompetenzmodell}
J.~Baumert and M.~Kunter, ``Das kompetenzmodell von coactiv,'' in \emph{Professionelle Kompetenz von Lehrkr{\"a}ften: Ergebnisse des Forschungsprogramms COACTIV}.\hskip 1em plus 0.5em minus 0.4em\relax Waxmann, 2011, pp. 29--53.

\bibitem{praetorius2018generic}
A.-K. Praetorius, E.~Klieme, B.~Herbert, and P.~Pinger, ``Generic dimensions of teaching quality: The german framework of three basic dimensions,'' \emph{Zdm}, vol.~50, pp. 407--426, 2018.

\bibitem{thiel2017}
F.~Thiel, D.~Ophardt, and V.~Barth, ``Staged videos zur störungsprävention und -intervention in der lehrerbildung - potenziale und entwicklung,'' \emph{Journal für LehrerInnenbildung}, 09 2017.

\bibitem{kervin2006classsim}
L.~Kervin, B.~Ferry, and L.~Carrington, ``Classsim: Preparing tomorrows teachers for classroom reality,'' in \emph{Society for Information Technology \& Teacher Education International Conference}.\hskip 1em plus 0.5em minus 0.4em\relax Association for the Advancement of Computing in Education (AACE), 2006, pp. 3204--3211.

\bibitem{lugrin2016breaking}
J.-L. Lugrin, M.~E. Latoschik, M.~Habel, D.~Roth, C.~Seufert, and S.~Grafe, ``Breaking bad behaviors: A new tool for learning classroom management using virtual reality,'' \emph{Frontiers in ICT}, vol.~3, p.~26, 2016.

\bibitem{hattie2009black}
J.~Hattie, ``The black box of tertiary assessment: An impending revolution,'' \emph{Tertiary assessment \& higher education student outcomes: Policy, practice \& research}, vol. 259, p. 275, 2009.

\bibitem{smith1993instructional}
P.~L. Smith and T.~J. Ragan, \emph{Instructional design}.\hskip 1em plus 0.5em minus 0.4em\relax Merrill Publishing Company., 1993.

\bibitem{lotz2015hattie}
M.~Lotz and F.~Lipowsky, ``Die hattie-studie und ihre bedeutung f{\"u}r den unterricht,'' \emph{Begabungen entwickeln und Kreativit{\"a}t f{\"o}rdern}, pp. 97--136, 2015.

\bibitem{Dieker2015TLETU}
\BIBentryALTinterwordspacing
L.~A. Dieker, M.~C. Hynes, C.~Hughes, S.~E. Hardin, and K.~M. Becht, ``Tle teachlive{\texttrademark}: Using technology to provide quality professional development in rural schools,'' \emph{Rural Special Education Quarterly}, vol.~34, pp. 11 -- 16, 2015. [Online]. Available: \url{https://api.semanticscholar.org/CorpusID:146586395}
\BIBentrySTDinterwordspacing

\bibitem{dieker2016mixed}
L.~Dieker, B.~Lignugaris-Kraft, M.~Hynes, and C.~Hughes, ``Mixed reality environments in teacher education: Development and future applications,'' \emph{Online in real time: Using Web 2.0 for distance education in rural special education}, pp. 122--131, 2016.

\bibitem{CONNOLLY2012661}
\BIBentryALTinterwordspacing
T.~M. Connolly, E.~A. Boyle, E.~MacArthur, T.~Hainey, and J.~M. Boyle, ``A systematic literature review of empirical evidence on computer games and serious games,'' \emph{Computers \& Education}, vol.~59, no.~2, pp. 661--686, 2012. [Online]. Available: \url{https://www.sciencedirect.com/science/article/pii/S0360131512000619}
\BIBentrySTDinterwordspacing

\bibitem{anuvziene2015structure}
I.~Anu{\v{z}}ien{\.e} \emph{et~al.}, ``The structure of socio-cultural competence (self) development,'' \emph{Profesinis rengimas: tyrimai ir realijos}, no.~26, pp. 94--105, 2015.

\bibitem{chang2009appraisal}
M.-L. Chang, ``An appraisal perspective of teacher burnout: Examining the emotional work of teachers,'' \emph{Educational psychology review}, vol.~21, pp. 193--218, 2009.

\bibitem{keller2014teachers}
M.~M. Keller, M.-L. Chang, E.~S. Becker, T.~Goetz, and A.~C. Frenzel, ``Teachers’ emotional experiences and exhaustion as predictors of emotional labor in the classroom: An experience sampling study,'' \emph{Frontiers in psychology}, vol.~5, p. 118409, 2014.

\bibitem{manstead2001social}
A.~S. Manstead, A.~H. Fischer \emph{et~al.}, ``Social appraisal: The social world as object of and influence on appraisal processes,'' \emph{Appraisal processes in emotion: Theory, methods, research}, pp. 221--232, 2001.

\bibitem{Gross2015}
J.~J. Gross, ``Emotion regulation: Current status and future prospects,'' \emph{Psychological Inquiry}, vol.~26, pp. 1--26, 1 2015.

\bibitem{ames2008taking}
D.~L. Ames, A.~C. Jenkins, M.~R. Banaji, and J.~P. Mitchell, ``Taking another person's perspective increases self-referential neural processing,'' \emph{Psychological science}, vol.~19, no.~7, pp. 642--644, 2008.

\bibitem{fonagy2006mechanisms}
P.~Fonagy and A.~W. Bateman, ``Mechanisms of change in mentalization-based treatment of bpd,'' \emph{Journal of clinical psychology}, vol.~62, no.~4, pp. 411--430, 2006.

\bibitem{scaffidi2016self}
C.~Scaffidi~Abbate, S.~Boca, and G.~H. Gendolla, ``Self-awareness, perspective-taking, and egocentrism,'' \emph{Self and Identity}, vol.~15, no.~4, pp. 371--380, 2016.

\bibitem{10.1145/1240624.1240696}
\BIBentryALTinterwordspacing
A.~Vasalou, A.~N. Joinson, and J.~Pitt, ``Constructing my online self: avatars that increase self-focused attention,'' in \emph{Proceedings of the SIGCHI Conference on Human Factors in Computing Systems}, ser. CHI '07.\hskip 1em plus 0.5em minus 0.4em\relax New York, NY, USA: Association for Computing Machinery, 2007, p. 445–448. [Online]. Available: \url{https://doi.org/10.1145/1240624.1240696}
\BIBentrySTDinterwordspacing

\bibitem{alves2023visual}
C.~Alves~da Silva, B.~Hilpert, C.~Bhuvaneshwara, P.~Gebhard, F.~Nunnari, and D.~Tsovaltzi, ``Visual similarity for socially interactive agents that support self-awareness,'' in \emph{Proceedings of the 23rd ACM International Conference on Intelligent Virtual Agents}, 2023, pp. 1--3.

\bibitem{https://doi.org/10.1017/S0048577201393198}
J.~J. Gross, ``Emotion regulation: Affective, cognitive, and social consequences,'' \emph{Psychophysiology}, vol.~39, no.~3, pp. 281--291, 2002.

\bibitem{AJZEN1991179}
\BIBentryALTinterwordspacing
I.~Ajzen, ``The theory of planned behavior,'' \emph{Organizational Behavior and Human Decision Processes}, vol.~50, no.~2, pp. 179--211, 1991, theories of Cognitive Self-Regulation. [Online]. Available: \url{https://www.sciencedirect.com/science/article/pii/074959789190020T}
\BIBentrySTDinterwordspacing

\bibitem{NEFF2007139}
\BIBentryALTinterwordspacing
K.~D. Neff, K.~L. Kirkpatrick, and S.~S. Rude, ``Self-compassion and adaptive psychological functioning,'' \emph{Journal of Research in Personality}, vol.~41, no.~1, pp. 139--154, 2007. [Online]. Available: \url{https://www.sciencedirect.com/science/article/pii/S0092656606000353}
\BIBentrySTDinterwordspacing

\bibitem{gilbert1989human}
\BIBentryALTinterwordspacing
P.~Gilbert, \emph{Human Nature and Suffering}, ser. Routledge revivals.\hskip 1em plus 0.5em minus 0.4em\relax Erlbaum, 1989. [Online]. Available: \url{https://books.google.de/books?id=A2mD-40vjG4C}
\BIBentrySTDinterwordspacing

\bibitem{mainhard2018student}
T.~Mainhard, S.~Oudman, L.~Hornstra, R.~J. Bosker, and T.~Goetz, ``Student emotions in class: The relative importance of teachers and their interpersonal relations with students,'' \emph{Learning and instruction}, vol.~53, pp. 109--119, 2018.

\bibitem{kunter2013development}
M.~Kunter, T.~Kleickmann, U.~Klusmann, and D.~Richter, ``The development of teachers’ professional competence,'' \emph{Cognitive activation in the mathematics classroom and professional competence of teachers: Results from the COACTIV project}, pp. 63--77, 2013.

\bibitem{dicke2015reducing}
T.~Dicke, J.~Elling, A.~Schmeck, and D.~Leutner, ``Reducing reality shock: The effects of classroom management skills training on beginning teachers,'' \emph{Teaching and teacher education}, vol.~48, pp. 1--12, 2015.

\bibitem{marzano2005handbook}
R.~J. Marzano, \emph{A handbook for classroom management that works}.\hskip 1em plus 0.5em minus 0.4em\relax ASCD, 2005.

\bibitem{hattie2008visible}
J.~Hattie, \emph{Visible learning: A synthesis of over 800 meta-analyses relating to achievement}.\hskip 1em plus 0.5em minus 0.4em\relax routledge, 2008.

\bibitem{Tausch2008PersonzentriertesVV}
\BIBentryALTinterwordspacing
R.~Tausch, ``Personzentriertes verhalten von lehrern in unterricht und erziehung,'' 2008. [Online]. Available: \url{https://api.semanticscholar.org/CorpusID:183354273}
\BIBentrySTDinterwordspacing

\bibitem{aspy1972investigation}
D.~N. Aspy and F.~N. Roebuck, ``An investigation of the relationship between student levels of cognitive functioning and the teacher’s classroom behavior,'' \emph{The Journal of Educational Research}, vol.~65, no.~8, pp. 365--368, 1972.

\bibitem{aspy1974humane}
------, ``From humane ideas to humane technology and back again many times.'' \emph{Education}, vol.~95, no.~2, 1974.

\bibitem{ryan2017self}
R.~M. Ryan and E.~L. Deci, \emph{Self-determination theory: Basic psychological needs in motivation, development, and wellness}.\hskip 1em plus 0.5em minus 0.4em\relax Guilford publications, 2017.

\bibitem{glock2020einstellungen}
S.~Glock, H.~Kleen, M.~Krischler, and I.~Pit-ten Cate, ``Die einstellungen von lehrpersonen gegen{\"u}ber sch{\"u}ler* innen ethnischer minorit{\"a}ten und sch{\"u}ler* innen mit sonderp{\"a}dagogischem f{\"o}rderbedarf: Ein forschungs{\"u}berblick,'' \emph{Stereotype in der Schule}, pp. 225--279, 2020.

\bibitem{nichols2004exploration}
J.~D. Nichols, ``An exploration of discipline and suspension data,'' \emph{Journal of Negro Education}, pp. 408--423, 2004.

\bibitem{rocque2011understanding}
M.~Rocque and R.~Paternoster, ``Understanding the antecedents of the" school-to-jail" link: The relationship between race and school discipline,'' \emph{The Journal of Criminal Law and Criminology}, pp. 633--665, 2011.

\bibitem{skiba2002color}
R.~J. Skiba, R.~S. Michael, A.~C. Nardo, and R.~L. Peterson, ``The color of discipline: Sources of racial and gender disproportionality in school punishment,'' \emph{The urban review}, vol.~34, pp. 317--342, 2002.

\bibitem{townsend2000disproportionate}
B.~L. Townsend, ``The disproportionate discipline of african american learners: Reducing school suspensions and expulsions,'' \emph{Exceptional children}, vol.~66, no.~3, pp. 381--391, 2000.

\bibitem{darnon2006mastery}
C.~Darnon, D.~Muller, S.~M. Schrager, N.~Pannuzzo, and F.~Butera, ``Mastery and performance goals predict epistemic and relational conflict regulation.'' \emph{Journal of educational psychology}, vol.~98, no.~4, p. 766, 2006.

\bibitem{schwarz2013konfliktmanagement}
\BIBentryALTinterwordspacing
G.~Schwarz, \emph{Konfliktmanagement: Konflikte erkennen, analysieren, l{\"o}sen}.\hskip 1em plus 0.5em minus 0.4em\relax Gabler Verlag, 2013. [Online]. Available: \url{https://books.google.de/books?id=N1LyBQAAQBAJ}
\BIBentrySTDinterwordspacing

\bibitem{rattay2011funf}
C.~Rattay and R.~Wensing, ``F{\"u}nf eskalationsstufen von unterrichtsst{\"o}rungen und m{\"o}gliche gegenmassnahmen,'' \emph{Unterrichtsst{\"o}rungen souver{\"a}n meistern--Das Praxisbuch. Profi-Tipps und Materialien aus der Lehrerfortbildung}, pp. 59--74, 2011.

\bibitem{nagy2012words}
W.~Nagy and D.~Townsend, ``Words as tools: Learning academic vocabulary as language acquisition,'' \emph{Reading research quarterly}, vol.~47, no.~1, pp. 91--108, 2012.

\bibitem{heppt2016verstandnis}
B.~Heppt, ``Verst{\"a}ndnis von bildungssprache bei kindern mit deutscher und nicht-deutscher familiensprache,'' 2016.

\bibitem{Gyurak2011}
A.~Gyurak, J.~J. Gross, and A.~Etkin, ``Explicit and implicit emotion regulation: A dual-process framework,'' pp. 400--412, 4 2011.

\bibitem{Duarte2016}
J.~Duarte, J.~Pinto-Gouveia, and B.~Cruz, ``Relationships between nurses' empathy, self-compassion and dimensions of professional quality of life: A cross-sectional study,'' \emph{International Journal of Nursing Studies}, vol.~60, pp. 1--11, 8 2016.

\bibitem{holt2005culture}
J.~L. Holt and C.~J. DeVore, ``Culture, gender, organizational role, and styles of conflict resolution: A meta-analysis,'' \emph{International Journal of Intercultural Relations}, vol.~29, no.~2, pp. 165--196, 2005.

\bibitem{hooi2012being}
R.~Hooi and H.~Cho, ``Being immersed: avatar similarity and self-awareness,'' in \emph{Proceedings of the 24th Australian Computer-Human Interaction Conference}, 2012, pp. 232--240.

\bibitem{barsalou1999perceptual}
L.~W. Barsalou, ``Perceptual symbol systems,'' \emph{Behavioral and brain sciences}, vol.~22, no.~4, pp. 577--660, 1999.

\bibitem{murray2017promoting}
D.~W. Murray and K.~Rosanbalm, ``Promoting self-regulation in adolescents and young adults: A practice brief. opre report 2015-82.'' \emph{Office of Planning, Research and Evaluation}, 2017.

\bibitem{gilbert2014origins}
P.~Gilbert, ``The origins and nature of compassion focused therapy,'' \emph{British journal of clinical psychology}, vol.~53, no.~1, pp. 6--41, 2014.

\bibitem{neff2003self}
K.~Neff, ``Self-compassion: An alternative conceptualization of a healthy attitude toward oneself,'' \emph{Self and identity}, vol.~2, no.~2, pp. 85--101, 2003.

\bibitem{sirois2015self}
F.~M. Sirois, ``A self-regulation resource model of self-compassion and health behavior intentions in emerging adults,'' \emph{Preventive medicine reports}, vol.~2, pp. 218--222, 2015.

\bibitem{prilop2019digital}
C.~N. Prilop, K.~E. Weber, and M.~Kleinknecht, ``How digital reflection and feedback environments contribute to pre-service teachers’ beliefs during a teaching practicum,'' \emph{Studies in Educational Evaluation}, vol.~62, pp. 158--170, 2019.

\bibitem{prilop2020effects}
------, ``Effects of digital video-based feedback environments on pre-service teachers’ feedback competence,'' \emph{Computers in Human Behavior}, vol. 102, pp. 120--131, 2020.

\bibitem{scheeler2004providing}
M.~C. Scheeler, K.~L. Ruhl, and J.~K. McAfee, ``Providing performance feedback to teachers: A review,'' \emph{Teacher education and special education}, vol.~27, no.~4, pp. 396--407, 2004.

\bibitem{ericsson2003development}
K.~A. Ericsson, J.~Starkes, and K.~Ericsson, ``Development of elite performance and deliberate practice,'' \emph{Expert performance in sports: Advances in research on sport expertise}, pp. 49--83, 2003.

\bibitem{protzel2015wertschatzung}
M.~Protzel, \emph{Wertsch{\"a}tzung und Pers{\"o}nlichkeitsentwicklung in der Grundschule: welchen Einfluss Lehrerhaltungen auf Kinder haben}.\hskip 1em plus 0.5em minus 0.4em\relax Optimus, 2015.

\bibitem{rahim2001managing}
\BIBentryALTinterwordspacing
M.~Rahim, \emph{Managing Conflict in Organizations, Third Edition}, ser. BusinessPro collection.\hskip 1em plus 0.5em minus 0.4em\relax Quorum Books, 2001. [Online]. Available: \url{https://books.google.de/books?id=0HjZAQAACAAJ}
\BIBentrySTDinterwordspacing

\bibitem{choi2023logs}
H.~Choi, P.~H. Winne, C.~Brooks, W.~Li, and K.~Shedden, ``Logs or self-reports? misalignment between behavioral trace data and surveys when modeling learner achievement goal orientation,'' in \emph{LAK23: 13th international learning analytics and knowledge conference}, 2023, pp. 11--21.

\bibitem{mehrabian1974basic}
A.~Mehrabian and J.~A. Russell, ``The basic emotional impact of environments,'' \emph{Perceptual and motor skills}, vol.~38, no.~1, pp. 283--301, 1974.

\bibitem{gebhard2012visual}
P.~Gebhard, G.~Mehlmann, and M.~Kipp, ``Visual scenemaker—a tool for authoring interactive virtual characters,'' \emph{Journal on Multimodal User Interfaces}, vol.~6, no.~1, p. 3–11, 2012.

\bibitem{schneeberger2019can}
T.~Schneeberger, M.~Scholtes, B.~Hilpert, M.~Langer, and P.~Gebhard, ``Can social agents elicit shame as humans do?'' in \emph{2019 8th International Conference on Affective Computing and Intelligent Interaction (ACII)}.\hskip 1em plus 0.5em minus 0.4em\relax IEEE, 2019, pp. 164--170.

\bibitem{schneeberger2019would}
T.~Schneeberger, S.~Ehrhardt, M.~S. Anglet, and P.~Gebhard, ``Would you follow my instructions if i was not human? examining obedience towards virtual agents,'' in \emph{2019 8th International Conference on Affective Computing and Intelligent Interaction (ACII)}.\hskip 1em plus 0.5em minus 0.4em\relax IEEE, 2019, pp. 1--7.

\bibitem{bhuvaneshwara2023mithos}
C.~Bhuvaneshwara, M.~Anglet, B.~Hilpert, L.~Chehayeb, A.-K. Meyer, D.~W. Don, D.~Tsovaltzi, P.~Gebhard, A.~Biermann, S.~Auchtor \emph{et~al.}, ``Mithos-mixed reality interactive teacher training system for conflict situations at school,'' in \emph{ISLS Annual Meeting 2023}, 2023, p.~42.

\bibitem{10.1145/2502081.2502223}
\BIBentryALTinterwordspacing
J.~Wagner, F.~Lingenfelser, T.~Baur, I.~Damian, F.~Kistler, and E.~Andr\'{e}, ``The social signal interpretation (ssi) framework: multimodal signal processing and recognition in real-time,'' ser. MM '13.\hskip 1em plus 0.5em minus 0.4em\relax New York, NY, USA: Association for Computing Machinery, 2013, p. 831–834. [Online]. Available: \url{https://doi.org/10.1145/2502081.2502223}
\BIBentrySTDinterwordspacing

\bibitem{mertes2024affecttoolbox}
S.~Mertes, D.~Schiller, M.~Dietz, E.~André, and F.~Lingenfelser, ``The affecttoolbox: Affect analysis for everyone,'' 2024.

\bibitem{rubenDiscourseAnalyser}
R.~García~Ucharima, ``{A Computational Model for DISCOURSE ANALYSIS, Applying Knowledge About SOCIAL NORMS and EMOTIONS},'' Master's thesis, Saarland University, 2023, retrieved from Saarland University Repository, Last accessed May 7, 2024.

\bibitem{poeschl2013german}
S.~Poeschl and N.~Doering, ``The german vr simulation realism scale-psychometric construction for virtual reality applications with virtual humans.'' \emph{Annual Review of Cybertherapy and Telemedicine}, vol.~11, pp. 33--37, 2013.

\bibitem{pertaub2002experiment}
D.-P. Pertaub, M.~Slater, and C.~Barker, ``An experiment on public speaking anxiety in response to three different types of virtual audience,'' \emph{Presence}, vol.~11, no.~1, pp. 68--78, 2002.

\bibitem{fitrianie202019}
S.~Fitrianie, M.~Bruijnes, D.~Richards, A.~B{\"o}nsch, and W.-P. Brinkman, ``The 19 unifying questionnaire constructs of artificial social agents: An iva community analysis,'' in \emph{Proceedings of the 20th ACM International Conference on Intelligent Virtual Agents}, 2020, pp. 1--8.

\bibitem{langer2022entwicklung}
W.~Langer, J.~Bruns, and J.~Erhorn, ``Entwicklung und validierung eines videobasierten testinstruments zur erfassung des noticing mit dem fokus auf anerkennungsprozesse im inklusiven sportunterricht,'' \emph{German Journal of Exercise and Sport Research}, vol.~52, no.~3, pp. 386--397, 2022.

\bibitem{seidel2010observer}
\BIBentryALTinterwordspacing
T.~Seidel, G.~Blomberg, and K.~St{\"u}rmer, ``„observer“ – validierung eines videobasierten instruments zur erfassung der professionellen wahrnehmung von unterricht. projekt observe,'' 2010. [Online]. Available: \url{https://api.semanticscholar.org/CorpusID:146458745}
\BIBentrySTDinterwordspacing

\bibitem{seufert2022classroom}
C.~Seufert, S.~Oberd{\"o}rfer, A.~Roth, S.~Grafe, J.-L. Lugrin, and M.~E. Latoschik, ``Classroom management competency enhancement for student teachers using a fully immersive virtual classroom,'' \emph{Computers \& Education}, vol. 179, p. 104410, 2022.

\bibitem{slater1999measuring}
M.~Slater \emph{et~al.}, ``Measuring presence: A response to the witmer and singer presence questionnaire,'' \emph{Presence: teleoperators and virtual environments}, vol.~8, no.~5, pp. 560--565, 1999.

\bibitem{downs2019polythetic}
E.~Downs, N.~D. Bowman, and J.~Banks, ``A polythetic model of player-avatar identification: Synthesizing multiple mechanisms.'' \emph{Psychology of popular media culture}, vol.~8, no.~3, p. 269, 2019.

\bibitem{van2010player}
J.~Van~Looy, C.~Courtois, and M.~De~Vocht, ``Player identification in online games: Validation of a scale for measuring identification in mmorpgs,'' in \emph{Proceedings of the 3rd International Conference on Fun and Games}, 2010, pp. 126--134.

\bibitem{seymour2021have}
M.~Seymour, L.~I. Yuan, A.~Dennis, K.~Riemer \emph{et~al.}, ``Have we crossed the uncanny valley? understanding affinity, trustworthiness, and preference for realistic digital humans in immersive environments,'' \emph{Journal of the Association for Information Systems}, vol.~22, no.~3, p.~9, 2021.

\end{thebibliography}

\end{document}